\documentclass[twocolumn, english, aps, prl, superscriptaddress]{revtex4-2}
\usepackage[T1]{fontenc}
\usepackage[latin9]{inputenc}
\usepackage{babel}
\usepackage{amsmath}
\usepackage{amssymb}
\usepackage{graphicx,xcolor}
\usepackage{esint}
\usepackage[unicode=true,pdfusetitle, bookmarks=true, bookmarksnumbered=false, bookmarksopen=false, breaklinks=false, pdfborder={0 0 1}, backref=false, colorlinks=false]
 {hyperref}
\usepackage{wasysym}
\usepackage{siunitx}
\usepackage[caption=false]{subfig}
\usepackage[normalem]{ulem}
\usepackage{url}
\usepackage{breakurl}


\newcommand{\Heff}[1]{\boldsymbol{H}_{\rm eff}\left[#1\right]}

\begin{document}
\title{Design of an optomagnonic crystal: towards optimal magnon-photon mode matching at the microscale}

\author{Jasmin Graf}
\affiliation{Max Planck Institute for the Science of Light, Staudtstra\ss{}e 2, 91058 Erlangen, Germany}
\affiliation{Department of Physics, University Erlangen-N{\"u}rnberg, Staudtstra\ss{}e 7, 91058 Erlangen, Germany}

\author{Sanchar Sharma}
\affiliation{Max Planck Institute for the Science of Light, Staudtstra\ss{}e 2, 91058 Erlangen, Germany}

\author{Hans Huebl}
\affiliation{Walther-Mei\ss{}ner-Institut, Bayerische Akademie der Wissenschaften, Walther-Meissner-Stra\ss{}e 8, 85748 Garching, Germany}
\affiliation{Physik-Department, Technische Universit{\"a}t M{\"u}nchen, James-Franck-Stra\ss{}e 1, 85748 Garching, Germany}
\affiliation{Munich Center for Quantum Science and Technology (MCQST), Schellingstra\ss{}e 4, 80799 M{\"u}nchen, Germany}

\author{Silvia {Viola Kusminskiy}}
\affiliation{Max Planck Institute for the Science of Light, Staudtstra\ss{}e 2, 91058 Erlangen, Germany}
\affiliation{Department of Physics, University Erlangen-N{\"u}rnberg, Staudtstra\ss{}e 7, 91058 Erlangen, Germany}

\begin{abstract}
We put forward the concept of an optomagnonic crystal: a periodically patterned structure at the microscale based on a magnetic dielectric, which can co-localize magnon and photon modes. The co-localization in small volumes can result in large values of the photon-magnon coupling at the single quanta level, which opens perspectives for quantum information processing and quantum conversion schemes with these systems. We study theoretically a simple geometry consisting of a one-dimensional array of holes with an abrupt defect, considering the ferrimagnet Yttrium Iron Garnet (YIG) as the basis material. We show that both magnon and photon modes can be localized at the defect, and use symmetry arguments to select an optimal pair of modes in order to maximize the coupling. We show that an optomagnonic coupling in the kHz range is achievable in this geometry, and discuss possible optimization routes in order to improve both coupling strengths and optical losses.

\end{abstract}
\maketitle

\section{Introduction} 
Progress in fundamental quantum physics has by now established a basis for developing new technologies in the fields of
information processing, secure communication, and quantum enhanced sensing. In order to perform these tasks, physical systems are needed which are capable of processing, storing, and communicating information in a quantum coherent manner and with a high fidelity. Similar to the classical realm, accomplishing this goal requires different degrees of freedom and efficient couplings between them, giving rise to {\it hybrid} systems. In this context, systems at the mesoscopic scale (with dimensions ranging from nanometers to microns) are specially interesting since their collective degrees of freedom can be tailored~\cite{hybrid_systems}. An important and successful example of these mesoscopic hybrid systems are optomechanical systems~\cite{optomechanics_review}, where light couples to mechanical motion. Seminal experiments in these systems have demonstrated extra-sensitive optical detection of small forces and displacements \cite{metcalfe_2014, zhou_2020, weber_2016, schreppler_2014, toros_2020, gosh_2020, de_groot_2019, tianran_2020}, manipulation and detection of mechanical motion in the quantum regime with light \cite{bergholm_2019, xu_2019, brawley_2016}, and the creation of nonclassical light and mechanical motion states \cite{bergholm_2019, baragiola_2018, yan_2016}.
\begin{figure}
\begin{centering}
\includegraphics[width=1\columnwidth]{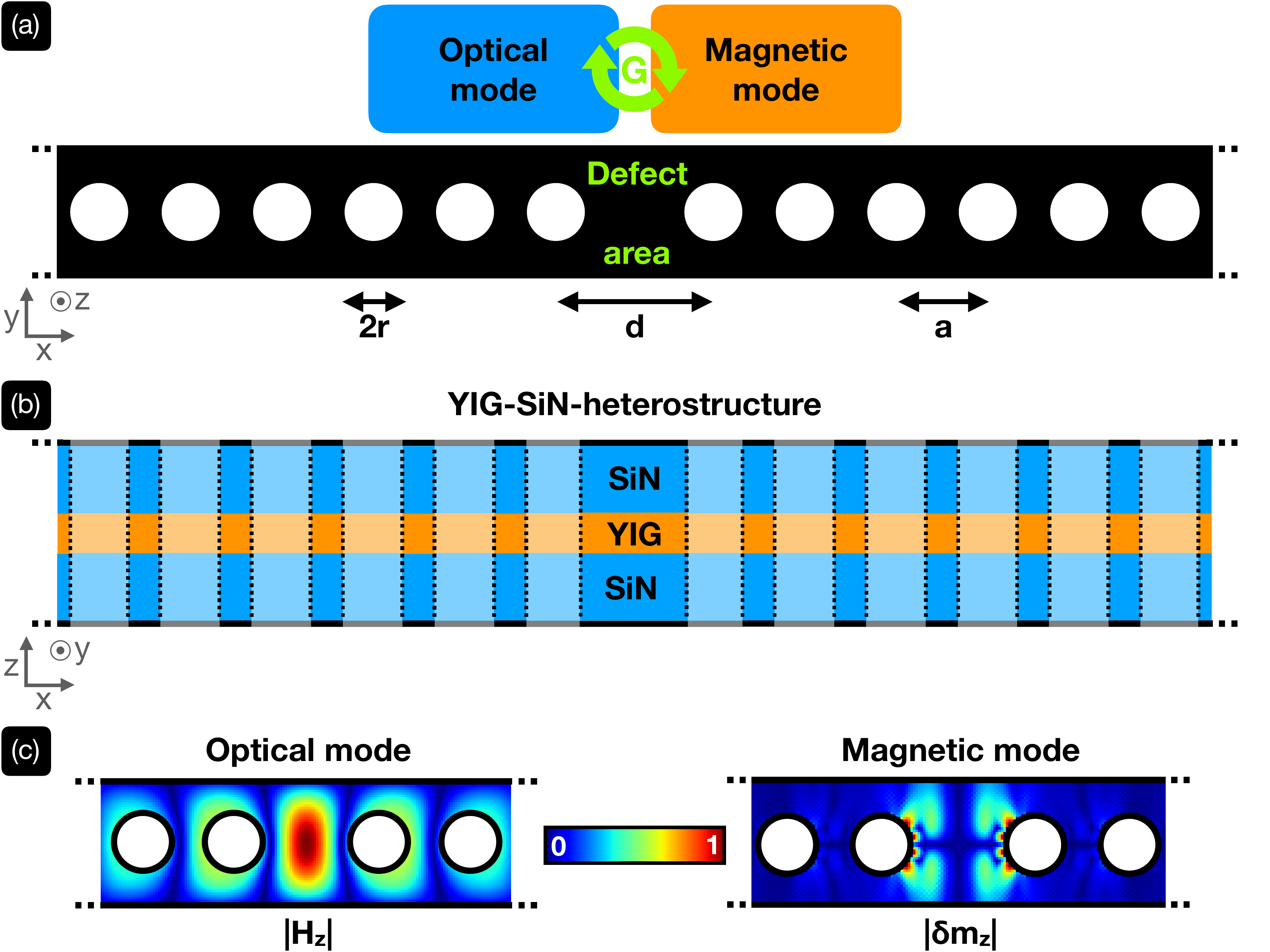}
\par\end{centering}
\caption{Investigated geometry: (a) Optomagnonic crystal with an abrupt defect at its center for localizing an optical and a magnon mode at the same spot in the defect area. (b) Optomagnonic crystal from the side representing a heterostructure. (c) Mode profiles of the localized optical and magnon mode discussed in the main text.\\
Note: all mode shape plots are normalized to their corresponding maximum value.}
 \label{figure_1}
\end{figure}

In the recent years, the family of hybrid quantum systems has been extended by incorporating magnetic materials, where the collective spin degree of freedom can be exploited. For example in spintronics~\cite{hybrid_systems_4},  information is carried by spins (as opposed to electrons) in order to remove Ohmic losses and to increase memory and processing capabilities~\cite{hybrid_system_4b, wimmer_2019}. An ultimate form of spintronics is the new field of quantum magnonics~\cite{tabuchi_quantum_2016, huebl_high_2013}, where superconducting quantum circuits couple, via microwave fields in a cavity, coherently to magnetic collective excitations (magnons)~\cite{hybrid_systems_6, tabuchi_coherent_2015}.
Such systems are promising for generating and characterizing non-classical quantum magnon states ~\cite{lachance-quirion_resolving_2016,lachance_single_magnon,tabuchi_quantum_2016, sharma_cat_states}, quantum thermometry protocols~\cite{potts_2020}, and for developing microwave-to-optical quantum transducers for quantum information processing~\cite{MtoO_magnons,hybrid_systems_6c}. The coherent coupling of magnons to optical photons has also been demonstrated in recent experiments ~\cite{optomagnonics_haigh, optomagnonics_osada, optomagnonics_zhang, optomagnonics_osada_3, optomagnonics_haigh_2}, in what have been denominated {\it optomagnonic systems}~\cite{optomagnonic_coupling_first,liu_optomagnonics_2016,optomagnonics_sharma,optomagnonic_new_2,optomagnonics_osada_2, hybrid_systems_6c,optomagnonics_review, optomagnonics_bittencourt}.

In current experiments exploring optomagnonics, the ferrimagnetic dielectric Yttrium Iron Garnet (YIG) is used as the magnetic element, since YIG presents low absorption and a large Faraday constant in the infrared ($\alpha = \SI{0.069}{cm^{-1}}$ and $\theta_{F} = \SI{240}{deg/cm}$ @ $\SI{1.2}{\micro\metre}$~\cite{faraday_2, yig_parameters_2, yig_wemple, optomagnonics_haigh, optomagnonics_osada}). The coupling between spins and optical photons is an second order processes involving spin-orbit coupling and it is generally small. This can be enhanced using a well polished sphere that acts as an optical cavity, trapping the photons by total internal reflection in order to effectively enhance the spin-photon coupling. The coupling, however, still remains too small for applications. This is due to the large size of the used YIG spheres (the coupling increases as the volume of the cavity decreases~\cite{optomagnonic_coupling_first}), with radius of the order of hundreds of microns, and, concomitantly, the large difference between the optical and the magnetic mode volume ($V_{\text{mag}}\gg V_{\text{opt}}$), by which most of the magnetic mode volume does not participate in the coupling.  This can be partially mitigated by making smaller cavities \cite{james_small_cavity, Alpanis_2020}, but care has to be taken both to obtain a good mode matching and to retain a good confinement of the optical mode in order to minimize radiation losses. Recent proposals have investigated one dimensional layered structures to this end~\cite{optomagnonic_new_1, optomagnonic_new_2}. 

In order to tackle these issues, we propose an optomagnonic array at the microscale, which acts simultaneously as a photonic crystal~\cite{photonic_crystals_molding_the_flow_of_light}, determining the optical properties of the structure, and as a magnonic crystal~\cite{magnonic_crystals_magnonics,  magnonic_crystals_review_and_prospects, magnonic_crystals_building_blocks_of_magnonics} with tailored magnetostatic modes. Our proposal is inspired in the success of this approach for optomechanical crystals, which can be designed such as to enhance the phonon-photon coupling by many orders of magnitude~\cite{opt_mech_crys_1D_1, opt_mech_crys_1D_2, opt_mech_crys_1D_3, opt_mech_crys_1D_4, opt_mech_crys_1D_5, opt_mech_crys_1D_6, opt_mech_crys_1D_7, opt_mech_crys_1D_8, opt_mech_crys_1D_9, opt_mech_crys_1D_10, opt_mech_crys_2D_1, opt_mech_crys_2D_2, opt_mech_crys_2D_3, opt_mech_crys_2D_4, opt_mech_crys_2D_5, opt_mech_crys_2D_7, opt_mech_crys_zip_1, opt_mech_crys_zip_2, opt_mech_crys_zip_3, opt_mech_crys_zip_4, opt_mech_crys_zip_5, opt_mech_crys_zip_6, opt_mech_crys_zip_7, opt_mech_crys_zip_8}. 
In our case, we use similar concepts in order to  design the coupling between photonic and magnonic modes. Although similar conceptually, magnetic materials present new challenges for the design, due to the complexity of the magnon modes.

Photonic crystals are the basis for many novel applications in quantum information, and are of high interest due to their ability to guide~\cite{guide_light_1, guide_light_2, guide_light_3, guide_light_4, guide_light_5} and confine~\cite{confine_light_1, confine_light_2, confine_light_3, confine_light_4, confine_light_5, confine_light_6} light, allowing for example to enhance non-linear optical interactions~\cite{nonlinear_light_1, nonlinear_light_2, nonlinear_light_3, nonlinear_light_4}. In turn, magnonic crystals can be designed to create reprogrammable magnetic band structures~\cite{reprogrammable_magnetic_band}, to act as band-pass or band-stop filters, or to create single-mode and bend waveguides~\cite{magnetic_filter_1, magnetic_filter_2, magnetic_filter_3, magnetic_filter_4}.
Additionally these crystals can be used for spin wave computing via logical gates~\cite{logical_gates_1, logical_gates_2, logical_gates_3}. An advantage of magnonic crystals is their scalability, their low energy consumption, and possibly faster operation rates~\cite{magnonic_crystals_review_and_prospects, logical_gates_3, magnonics_advantages_2}. Together with the state of the art in optomagnonics detailed above, this provides a great incentive to explore the possibility of an {\it{optomagnonic}} crystal, combining both photonic and magnetic degrees of freedom. 

Specifically, we consider an optomagnonic crystal consisting of a dielectric magnetic slab  (YIG in our simulations) with a periodic array  of  holes along the slab and with an abrupt defect in the middle. The repeated holes at each side of the defect act as a Bragg mirror for both optical and magnetic modes, localizing them in the region of the defect (see Fig.~\ref{figure_1}(a)+(c)). We show that this structure can co-localize photonic and magnonic modes, and explore how the symmetry of the modes can be used to optimize their coupling. We find that coupling rates in the range of kHz are achievable in these structures, and that optimization of the geometry can lead to higher coupling values, indicating the promise of this approach. Further optimization is nevertheless needed to improve the decay rates, in particular the optical quality factor is low compared to the state of the art in non-magnetic structures (where Silicon is used as the dielectric).

This manuscript is organized as follows. In Sec.~\ref{optomagnonic_coupling_sec} we derive the general expression for the coupling of magnons to optical photons and discuss the normalization of the modes required to find the photon-magnon coupling at the single quanta level, denominated {\it{optomagnonic coupling}}. The remaining sections refer to the numerical method for evaluating this coupling. For our simulations we choose YIG as the magnetic material, in line with the material of choice in experiments. In Sec.~\ref{photonic_crystal_sec} we discuss the properties of the proposed structure as a photonic crystal. In Sec.~\ref{magnonic_crystal_sec}, in turn, we investigate its properties as a magnonic crystal. Sec.~\ref{optomagnonic_crystal_sec} combines the results in order to numerically evaluate the optomagnonic coupling for appropriately chosen confined modes. For concreteness, we focus on the coupling between one single magnon mode and one single optical mode. Sec.~\ref{optimization_sec} is devoted to a discussion on how the structure can be optimized and presents results for an optimized geometry. The conclusions and an outlook are presented in Sec.~\ref{conclusion_sec}. The supplementary material contains further details of the analytic calculations and of the simulations.

\section{Optomagnonic coupling} \label{optomagnonic_coupling_sec}
In this section we derive the theoretical expression for the coupling rate between magnons and photons.
The instantaneous electromagnetic energy is~\cite{energy}
\begin{equation}
E_{\text{em}}^{\text{tot}} = \frac{1}{2} \int \text{d} \boldsymbol{r} ~ \big[ \boldsymbol{\mathcal{D}}(\boldsymbol{r},t) \cdot \boldsymbol{\mathcal{E}}(\boldsymbol{r},t) + \boldsymbol{\mathcal{B}}(\boldsymbol{r},t) \cdot \boldsymbol{\mathcal{H}}(\boldsymbol{r},t) \big]
\label{em_energy}
\end{equation}
with $\boldsymbol{\mathcal{D}}$ the displacement field, $\boldsymbol{\mathcal{E}}$ the electric field, $\boldsymbol{\mathcal{B}}$ the magnetic induction, and $\boldsymbol{\mathcal{H}}$ the magnetic field.
In complex representation $\boldsymbol{\mathcal{D}} = (\boldsymbol{D} + \boldsymbol{D}^*)/2$ and $\boldsymbol{\mathcal{E}} = (\boldsymbol{E} + \boldsymbol{E}^*)/2$, and similar for the magnetic induction and field. The effect of the magnetization $\boldsymbol{\mathcal{M}}$ is to modify the displacement field as $\boldsymbol{D}(\boldsymbol{r},t) = \bar{\varepsilon} [ \boldsymbol{\mathcal{M}}(\boldsymbol{r},t)] \cdot \boldsymbol{E}(\boldsymbol{r},t)$ where the components of the permittivity tensor $\bar{\varepsilon}$ are~\cite{cotton_mouton,light_magnetics_2}
\begin{equation}
\varepsilon_{ij}(\boldsymbol{\mathcal{M}}) = \varepsilon_0 \left( \varepsilon_{\text{r}} \delta_{ij} - i f_{\text{F}} \sum_k \epsilon_{ijk} \mathcal{M}_k + f_{\text{C}} \mathcal{M}_i \mathcal{M}_j \right),
\label{EpsFaraday}
\end{equation}
with $\varepsilon_0$ the vacuum permittivity, $\varepsilon_{\text{r}}$ the relative permittivity, $\epsilon_{ijk}$ the Levi-Cevita tensor, and $\{f_{\text{F}},f_{\text{C}}\}$ material dependent magneto-optical constants. 
At optical frequencies the second term in Eq.~\eqref{em_energy} can be neglected~\cite{light_magnetics_1, light_magnetics_2}, being smaller than the first by the fine structure constant squared, and the permeability of the material can be set to the vacuum permeability $\mu_0$.
The magneto-optical constants can be related to the Faraday rotation $\theta_{\text{F}}$ and the Cotton-Mouton ellipticity $\theta_{\text{C}}$ per unit length as $\theta_{\text{F}} = \omega/(2c \sqrt{\varepsilon_{\text{r}}}) f_{\text{F}} \,M_{\text{s}}$ and $\theta_{\text{C}} = \omega/(2c \sqrt{\varepsilon_{\text{r}}}) f_{\text{C}} \,M_{\text{s}}^2$, with $c$ the speed of light in vacuum and $M_{\text{s}}$ the saturation magnetization. 

We are interested in how light couples to the {\it{fluctuations}} of the magnetization around the static ground state. We consider norm-preserving small fluctuations, 
\begin{equation} 
    \boldsymbol{\cal M}(\boldsymbol{r},t) = \boldsymbol{M}_0(\boldsymbol{r}) \sqrt{1- \left|\frac{\delta\boldsymbol{M}(\boldsymbol{r},t)}{M_{\text{s}}}\right|^2} + \delta\boldsymbol{M}(\boldsymbol{r},t), 
\end{equation}
where the ground state satisfies $\boldsymbol{M}_0\cdot\boldsymbol{M}_0 = M_{\text{s}}^2$ and the fluctuations are perpendicular to local equilibrium magnetization $\delta\boldsymbol{M}\cdot\boldsymbol{M}_0 = 0$. In complex notation $\delta \boldsymbol{M}=[\boldsymbol{M} + \boldsymbol{M}^*]/2$.
The correction to the electromagnetic energy stemming from the interaction between the light field and the magnetization can be rewritten as 
\begin{equation}
E_{\text{em}} = \frac{1}{8} \int \text{d} \boldsymbol{r} ~ \big[ \boldsymbol{E}(\boldsymbol{r},t) \cdot \boldsymbol{D}^*(\boldsymbol{r},t) + \boldsymbol{E}^*(\boldsymbol{r},t) \cdot \boldsymbol{D}(\boldsymbol{r},t) \big],
\end{equation}
ignoring $\boldsymbol{E}(\boldsymbol{r},t) \cdot \boldsymbol{D}(\boldsymbol{r},t)$ and $\boldsymbol{E}^*(\boldsymbol{r},t) \cdot \boldsymbol{D}^*(\boldsymbol{r},t)$ in the rotating wave approximation.
Inserting the relation between the displacement and the electric field along with the permittivity in Eq.~\eqref{EpsFaraday} gives $E_{\text{em}} = E_{\text{em}}^{\text{F}}+E_{\text{em}}^{\text{C}}$ where
\begin{equation}
E_{\text{em}}^{\text{F}} = \frac{\varepsilon_0 f_{\text{F}}}{8} \int \text{d} \boldsymbol{r} \left[ i\left(\boldsymbol{E}^*\times \boldsymbol{E}\right)\cdot \boldsymbol{M} + \text{h.c.}\right] 
\end{equation}
is the Faraday contribution and
\begin{equation}
	E_{\rm em}^{\text{C}} = \frac{\epsilon_0 f_{\text{C}}}{8} \int \text{d}\boldsymbol{r} \left[ \boldsymbol{E}^*\cdot\left( \boldsymbol{M}\boldsymbol{M}_0 + \boldsymbol{M}_0 \boldsymbol{M} \right)\cdot\boldsymbol{E} + \text{h.c.} \right]
\end{equation}
the Cotton-Mouton one. We have used the dyadic notation and neglected all terms that represent a constant energy shift or that are higher order in $\delta\boldsymbol{M}$.

Quantizing this expression leads to the optomagnonic coupling Hamiltonian.
By assuming that the magnetic material acts as an optical cavity, the electric field of the light can be quantized by using the annihilation (creation) operator $\hat{a}_{\beta}^{(\dagger)}$ of one photon
\begin{equation}
\boldsymbol{E}(\boldsymbol{r},t) \rightarrow 2 i \sum_{\alpha} \boldsymbol{E}_{\alpha} (\boldsymbol{r}) \, \hat{a}_{\alpha}(t),
\end{equation}
with $\boldsymbol{E}_{\alpha}^{(*)}$ the mode shape, and $\alpha$ the mode index.
We note that we identified $\boldsymbol{E}(\boldsymbol{r},t)$ with $2 \, \boldsymbol{E}^+(\boldsymbol{r},t)$ from the well known quantization expression of the electric field~\cite{dirac}
\begin{equation}
\begin{split}
\boldsymbol{\mathcal{{E}}}(\boldsymbol{r},t) &= \boldsymbol{E}^+(\boldsymbol{r},t) + \boldsymbol{E}^-(\boldsymbol{r},t) \\[0,2cm] 
&= i \sum_{\alpha}  \Big[ \boldsymbol{E}_{\alpha} (\boldsymbol{r}) \, \hat{a}_{\alpha}(t) - \boldsymbol{E}_{\alpha}^* (\boldsymbol{r}) \, \hat{a}_{\alpha}^{\dagger}(t) \Big].
\end{split}
\end{equation}
In order to find the coupling per photon, we normalize the electromagnetic field amplitude to one photon over the electromagnetic vacuum~\cite{optomechanics_book}
\begin{equation}
\frac{\hbar \omega_{\alpha}}{2 \varepsilon_0} =  \int \text{d} \boldsymbol{r} \,  \varepsilon_{\text{r}}(\boldsymbol{r}) |\boldsymbol{E}_{\alpha}(\boldsymbol{r})|^2.
\label{norm_optics}
\end{equation}

The spin waves can be quantized as
\begin{equation}
\boldsymbol{M}(\boldsymbol{r},t) \rightarrow 2M_{\text{s}} \sum_{\gamma} \delta \boldsymbol{m}_{\gamma} (\boldsymbol{r}) \, \hat{b}_{\gamma}(t)  ,
\end{equation}
where $\hat{b}_{\gamma}^{(\dagger)}$ annihilates (creates) one magnon, $ \delta \boldsymbol{m}_{\gamma}^{(*)}$ is the mode shape, and $\gamma$ the mode index.
We note that as in the optical case we identified $\boldsymbol{M}(\boldsymbol{r},t)$ with $2 \, \boldsymbol{M}^+(\boldsymbol{r},t)$ from the magnetic quantization expression 
\begin{equation}
\begin{split}
\delta\boldsymbol{M}(\boldsymbol{r},t) &= 
 \boldsymbol{M}^+(\boldsymbol{r},t) + \boldsymbol{M}^-(\boldsymbol{r},t) \\[0,2cm] 
&= M_{\text{s}} \sum_{\gamma}  \Big[ \delta \boldsymbol{m}_{\gamma} (\boldsymbol{r}) \, \hat{b}_{\gamma}(t) + \delta \boldsymbol{m}_{\gamma}^* (\boldsymbol{r}) \, \hat{b}_{\gamma}^{\dagger}(t) \Big].
\end{split}
\end{equation}
In order to normalize the amplitude of the magnetic fluctuations to one magnon, we use the following expression derived in the supplementary material (see appendix A)
\begin{equation}
\frac{g \mu_{\text{B}}}{M_{\text{s}}} = \int \text{d} \boldsymbol{r} \, i \boldsymbol{m}_0 \cdot \big[ \delta \boldsymbol{m}_{\gamma}^* \times \delta \boldsymbol{m}_{\gamma}\big]
\label{norm_magnetics}
\end{equation}
with $g$ the g-factor, $\mu_{\text{B}}$ the Bohr magneton, and $\boldsymbol{m}_0=\boldsymbol{M}_0/M_{\text{s}}$.
This expression is valid for arbitrary magnetic textures and it is consistent with the normalization derived previously for a uniform ground state \cite{mills_quantum_magnons}.

The quantized optomagnonic energy, neglecting the constant energy shifts, leads to the coupling Hamiltonian
\begin{equation}
\hat{H}_{om} = \hbar \sum_{\alpha \beta \gamma} \Big[ G_{\alpha \beta \gamma} \, \hat{a}_{\alpha}^{\dagger} \hat{a}_{\beta} \, \hat{b}_{\gamma}  + \text{h.c.} \Big]\label{eq:Hom}
\end{equation}
with the coupling constant $G_{\alpha\beta\gamma} = G^{\text{F}}_{\alpha\beta\gamma} + G^{\text{C}}_{\alpha\beta\gamma}$, where
\begin{align}
    G^{\text{F}}_{\alpha\beta\gamma} &= -i\frac{\varepsilon_0\varepsilon_{\text{r}}}{\hbar}\frac{\theta_{\text{F}}\lambda_{n}}{\pi}  \int \text{d} \boldsymbol{r}  ~ \big[\boldsymbol{E}_{\alpha}^* \times \boldsymbol{E}_{\beta} \big] \cdot \delta \boldsymbol{m}_{\gamma}, \label{optomagnonic_coupling} \\
    G^{\text{C}}_{\alpha\beta\gamma} &= \frac{\varepsilon_0\varepsilon_{\text{r}}}{\hbar}\frac{\theta_{\text{C}}\lambda_{n}}{\pi} \int \text{d}\boldsymbol{r} ~ \left[ \boldsymbol{m}_0 \cdot \boldsymbol{E}_{\beta} \right] \left[ \boldsymbol{E}_{\alpha}^* \cdot \delta \boldsymbol{m}_{\gamma} \right] \nonumber \\
    &+\frac{\varepsilon_0\varepsilon_{\text{r}}}{\hbar}\frac{\theta_{\text{C}}\lambda_{n}}{\pi} \int \text{d}\boldsymbol{r} ~ \left[ \boldsymbol{m}_0 \cdot \boldsymbol{E}_{\alpha}^* \right] \left[ \boldsymbol{E}_{\beta} \cdot \delta \boldsymbol{m}_{\gamma} \right] 
    \label{CM_coupling}
\end{align}
are, respectively, the Faraday and Cotton-Mouton components of the optomagnonic coupling constant,  being $\lambda_{n}$ the light wavelength inside the material.

The coupling between optical photons and magnons, as can be seen from Eq.~\ref{eq:Hom}, involves a three-particle process in which a magnon is created or annihilated by a two-photon scattering process. This is an example of \emph{parametric} coupling, and reflects the frequency mismatch between the excitations. The coupling can be enabled by a triple-resonance, where the frequency of the magnon matches the frequency difference between two photonic modes~\cite{optomagnonics_haigh, optomagnonics_osada,optomagnonics_zhang, liu_optomagnonics_2016}, or, in the case of scattering with a single photon mode, by an external driving laser at the right detuning~\cite{optomagnonic_coupling_first}. If the laser is red (blue) detuned by the magnon frequency, implying a lower (higher) driving frequency than the photon resonance, it will annihilate (create) magnons. In the red detuned regime this can be used, for example, to actively cool the magnon mode to its ground state~\cite{optomechanics_review, optomagnonics_bittencourt, sharma_optical_2018}.

In this work we focus on the coupling between a given magnon mode and a given optical mode hosted by the 1D optomagnonic crystal shown in Fig.~\ref{figure_1}(a)+(b). 
Hence we set $\alpha = \beta$ and drop the indices in the following.
For concreteness, we focus on $G^{\text{F}}$ as the analysis for $G^{\text{C}}$ is analogous. $G^{\text{F}}$ is proportional to the overlap of the magnon's spatial distribution with the electric component of the optical spin density defined as~\cite{silvia_book}
\begin{equation}
    \boldsymbol{S}_{\text{opt}} (\boldsymbol{r}) = \frac{\varepsilon_0}{2 i \, \omega_{\text{opt}}} \big[\boldsymbol{E}^* \times \boldsymbol{E} \big].
\end{equation}
The optical spin density is finite only for fields with certain degree of circular polarization, and points perpendicular to the plane of polarization. 

\section{Photonic crystal} 
\label{photonic_crystal_sec}
Photonic crystals are engineered structures which, by proper shape design, can confine light to a specific region. These are formed by low-loss media exhibiting a periodic dielectric function $\varepsilon(\boldsymbol{r})$, with a discrete translational symmetry $\varepsilon(\boldsymbol{r}) = \varepsilon(\boldsymbol{r} + \boldsymbol{R})$ for any $\boldsymbol{R} = n \boldsymbol{a}$ with $n$ an integer and $\boldsymbol{a}$ the lattice constant given by the imposed periodicity.

\begin{figure}
\begin{centering}
\includegraphics[width=1\columnwidth]{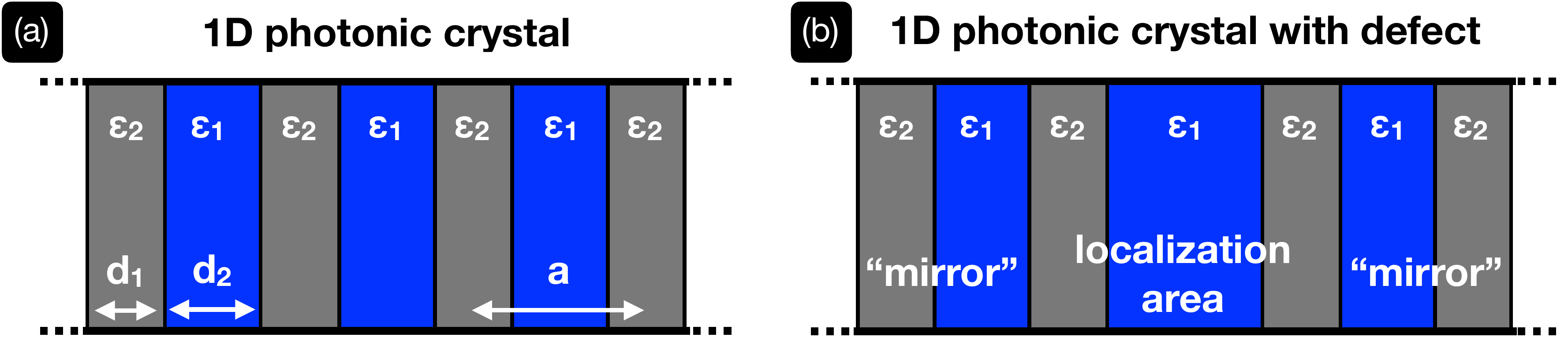}
\par\end{centering}
\caption{General structure of a 1D photonic crystal and mode localization at a defect: (a) 1D photonic crystal consisting of periodic layers alternated by the lattice constant $a$ with different dielectric constants $\varepsilon_1 > \varepsilon_2$ and widths $d_1$ and $d_2$. (b) A defect breaks the symmetry and can pull a band-edge mode into the photonic band gap. Since a mode in the band gap cannot propagate into the structure, the light is Bragg-reflected and is thus localized (see e.g.~\cite{photonic_crystals_molding_the_flow_of_light}).}
 \label{figure_2}
\end{figure}

Photonic band gaps arise at the edges of the Brillouin zone (BZ) $k = \pi/a$ due to the periodicity imposed by the susceptibility of the crystal on the electric field, with wavelength $\lambda = 2a$ (corresponding to the edge of the BZ). For example, in a 1D photonic crystal (see Fig.~\ref{figure_2}(a)) the symmetry of the unit cell around its center implies that the nodes of the standing light wave must be centered either at each low-$\varepsilon$ layer or at each high-$\varepsilon$ layer.
The latter necessarily has lower energy than the former, resulting in a band gap. The position of the photonic band gap is given by the mid-gap frequency at the BZ edge. 
In the case of two materials with refractive indices $n_1$ and $n_2$ and thicknesses $d_1$ and $d_2 = a - d_1$, the normal incidence gap is maximized for $n_1 d_1 = n_2 d_2$. 
In this case the mid-gap frequency is given by~\cite{photonic_crystals_molding_the_flow_of_light}
\begin{equation}
\omega_{\text{mg}} = \frac{n_1 + n_2}{4 n_1 n_2} \cdot \frac{2 \pi c}{a}
\label{mid-gap-frequency}
\end{equation}
with $n_1 = \sqrt{\varepsilon_1}$ , $n_2 = \sqrt{\varepsilon_2}$. 
The corresponding vacuum wavelength $\lambda_{\text{mg}} = (2 \pi c)/\omega_{\text{mg}}$ thereby satisfies the relations $\lambda_{\text{mg}} /n_1 = 4 d_1$ and $\lambda_{\text{mg}} /n_2 = 4 d_2$ meaning that the individual layers are a quarter-wavelength thick. 

An input at frequencies within the photonic band gap is reflected entirely except for an exponentially decaying tail inside the crystal.
Thus, two of such crystals can be used to create a Fabry-Perot like cavity.
More concretely, as shown in Fig.~\ref{figure_2}(b), a defect in the form of a layer with a different width breaking the symmetry of the crystal may permit localized modes in the band gap by consecutive reflection on both sides.
Since the light is localized in a finite region, the modes are quantized into discrete frequencies.
We note that the degree of localization is the largest for modes with frequencies near the center of the gap~\cite{photonic_crystals_molding_the_flow_of_light}.

For our purposes, we consider a geometry in which the permittivity can take two distinct values, attained by holes carved into a dielectric slab (see Fig.~\ref{figure_1}(a)). The typical material used for photonic crystals is silicon due to its high refractive index at optical frequencies, $\varepsilon_{\text{r}}=12$. We use instead YIG for our study,  which is a dielectric magnetic material transparent in the infrared range with $\varepsilon_{\text{r}}=5$ \cite{yig_optical_absorption}. The lower dielectric constant reduces the confinement of the optical modes along the height of the slab, which is reflected in low optical quality factors as discussed below.
This structure is a 1D photonic crystal, periodic in one direction (chosen to be the $\hat{x}$-direction), with a band gap along this direction and which confines light through index guiding~\cite{photonic_crystals_molding_the_flow_of_light} (a generalization of total internal reflection) in the remaining directions.
In order to localize an optical mode in this structure we create a defect by increasing the spacing between the two middle holes, which pulls a mode into the band gap. 
We note that due to the (discrete) periodicity, the crystal only possesses an incomplete band gap and the localized mode can scatter to air modes~\cite{photonic_crystals_molding_the_flow_of_light}.

We search for a localized mode in the infrared frequency range where YIG is transparent and presents low absorption~\cite{yig_parameters_2,yig_absorption}). 
Thus, the geometrical parameters of the crystal need to be chosen in such a way that the band gap lies in the desired frequency range.
We choose a lattice constant of $a=\SI{450}{\nano\metre}$ which gives a mid-gap frequency of $\omega_{\text{mg}} = 2\pi\times\SI{240}{\tera\hertz}$ (corresponding to $\lambda \approx \SI{1250}{\nano\metre}$), using Eq.~\eqref{mid-gap-frequency} with refractive indices of YIG $n_1 = n_{\rm{YIG}} = \sqrt{5}$ and of air $n_2 = n_{\text{air}} = 1$. Note that we choose a lattice constant that allows us to work in the transparency frequency range for the optics, and which at the same time is small enough in order to reduce the computational cost of the micromagnetic simulations of the corresponding magnonic crystal in the next section. 
Using the relation $d_{\text{air}} n_{\text{air}} = d_{\rm{YIG}} n_{\rm{YIG}}$ for a maximized normal incidence gap we find the optimal radius of the air holes as
\begin{equation}
r_{\text{air}} = \frac{n_{\rm{YIG}}}{n_{\rm{YIG}} + 1} \cdot \frac{a}{2}
\end{equation}
with $d_{\text{air}} = 2r_{\text{air}} = a - d_{\rm{YIG}}$, from which we obtain $r_{\text{air}} = \SI{155.25}{\nano\metre}$.
In order to find the mode with the least losses, the defect width is optimized in order to localize the desired mode most effectively to the defect.  
We find the optimal defect size, defined as the center-to-center distance between the two bounding holes (see Fig.~\ref{figure_1}(a)), to be $d = \SI{731}{\nano\metre}$,  obtained  by  evaluating the transmission spectra as a function of the defect size (we used the electromagnetic simulation tool MEEP to this end~\cite{MEEP}).
In order to get a good quality factor of the localized mode we need to insert as many air holes as possible.
For creating a compromise between short computational evaluation time (especially important for the magnetic simulations discussed later) and a good quality factor we chose $N=12$ holes at each side of the defect.
Therefore the investigated crystal is in total $l = \SI{11.75}{\micro\metre}$ long.
The overall width of the wave guide is $w = \SI{600}{\nano\metre}$ and its height is $h = \SI{60}{\nano\metre}$, again to keep the magnetic simulations (which we detail in the next section) feasible. Such a thin slab will not be good at confining the optical modes along its height, since it is much smaller than the light wavelength in the material. In order to increase the optical quality factor without influencing the magnetics we sandwich the crystal with two $\text{Si}_3\text{N}_4$ layers (see Fig.~\ref{figure_1}(b)) with a height of $h_{\text{Si}_3\text{N}_4}= \SI{200}{\nano\metre}$ as proposed in~\cite{my_paper}. $\text{Si}_3\text{N}_4$ has an index of refraction similar to YIG ($n_{\text{Si}_3\text{N}_4} = \sqrt{4}$), so that the combined structure acts approximately as a single cavity for the light and its height is roughly half a wavelength, enough to provide a reasonable confinement. The presented simulations include these two extra layers. 

\begin{figure}
\begin{centering}
\includegraphics[width=1\columnwidth]{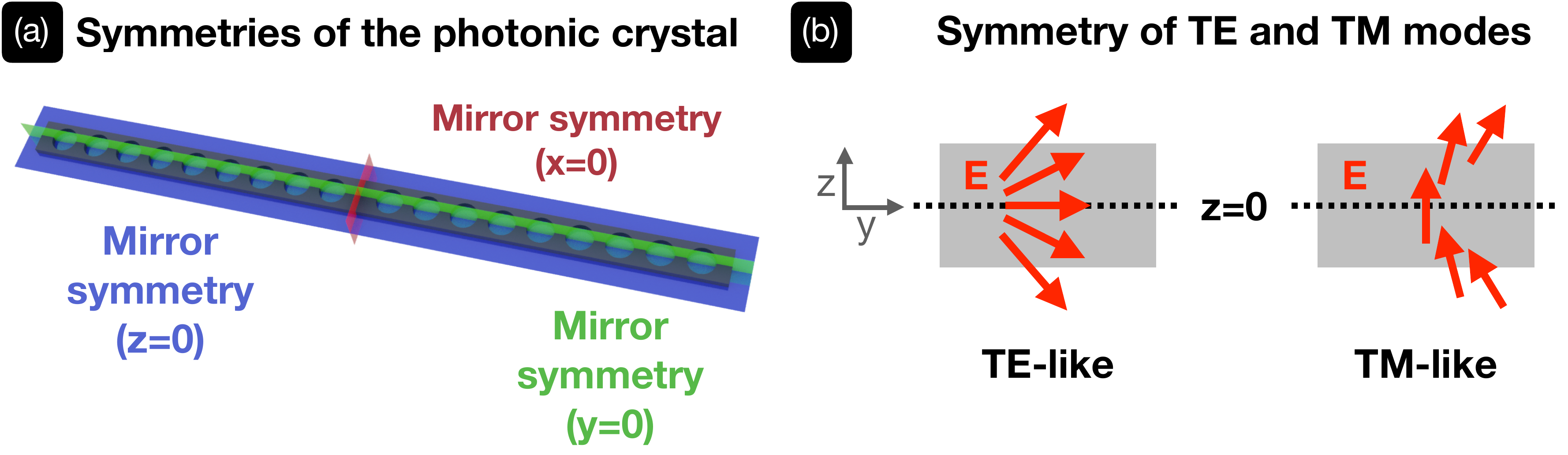}
\par\end{centering}
\caption{Symmetries of the optical modes in a periodic waveguide: (a) Symmetry planes of the investigated 1D photonic crystal shown in Fig.~\ref{figure_1}(a). (b) Symmetry of a transverse electric (TE)-like and a transverse magnetic (TM)-like optical mode in a thin 3D structure. The red arrows indicate the electric field vector $\boldsymbol{E}$ which for $z=0$ (middle of the crystal along the height) lie in plane for TE-like modes and point out of plane for TM-like modes. For $z \neq 0$ this is not fulfilled anymore (see e.g.~\cite{photonic_crystals_molding_the_flow_of_light}).}
 \label{figure_3}
\end{figure}

We now turn to categorizing the photonic modes of the crystal by  using its three mirror symmetry planes (see Fig.~\ref{figure_3}(a)). This imposes several restrictions on the mode shape and the mode polarization. We define the three mirror symmetry operations 
\begin{equation}
\begin{split}
\hat{\sigma}_z^{\text{E}} \, \boldsymbol{E}(\boldsymbol{r}) &= \left( \begin{array}{c}
E_x (x,y,-z) \\ 
E_y (x,y,-z) \\ 
-E_z (x,y,-z)
\end{array}  \right), 
\\[0,2cm]
\hat{\sigma}_y^{\text{E}} \, \boldsymbol{E}(\boldsymbol{r}) &= \left( \begin{array}{c}
E_x (x,-y,z) \\ 
-E_y (x,-y,z) \\ 
E_z (x,-y,z)
\end{array}  \right), 
\\[0,2cm]
\hat{\sigma}_x^{\text{E}} \, \boldsymbol{E}(\boldsymbol{r}) &= \left( \begin{array}{c}
-E_x (-x,y,z) \\ 
E_y (-x,y,z) \\ 
E_z (-x,y,z)
\end{array}  \right) .
\end{split}
\label{symmetry_defs}
\end{equation}
In the following we restrict the discussion to transverse electric modes with an in-plane electric field, which are the modes of interest for the magnetic configuration we consider, as it will be clear from the next section. 
We note that structures made of a  high-$\varepsilon$ material with air holes favour a band gap for transverse electric modes~\cite{photonic_crystals_molding_the_flow_of_light}, which is advantageous for our purposes. Unlike in 2D, in three dimensions we cannot generally distinguish between transverse electric (TE) and transverse magnetic (TM) modes. However, provided that the crystal has a mirror symmetry along its height (under $\hat{\sigma}_z^{\text{E}}$), and that its thickness is smaller than the mode wavelength, the fields are mostly polarized in TE-like and TM-like modes~\cite{photonic_crystals_molding_the_flow_of_light} (see Fig.~\ref{figure_3}(b)).
Since a TE-like mode has a non-zero electric field in the plane of the crystal ($xy$-plane), both $E_x$ and $E_y$ cannot be odd as a function of $z$ (see Fig.~\ref{figure_3}(b)). From Eq.~(\ref{symmetry_defs}), this implies that the mode must be even under $\hat{\sigma}_z^{\text{E}}$. Similar symmetry considerations~\cite{photonic_crystals_molding_the_flow_of_light} show that a TE-like mode must satisfy 
\begin{equation}
\begin{split}
    \hat{\sigma}_z^{\text{E}} \boldsymbol{E}(\boldsymbol{r}) &= \boldsymbol{E}(\boldsymbol{r}),\\[0,2cm] \hat{\sigma}_y^{\text{E}} \boldsymbol{E}(\boldsymbol{r}) &= -\boldsymbol{E}(\boldsymbol{r}),\\[0,2cm] \hat{\sigma}_x^{\text{E}} \boldsymbol{E}(\boldsymbol{r}) &= -\boldsymbol{E}(\boldsymbol{r}). \label{TE-like_mode}
\end{split}
\end{equation}
\begin{figure*}[t!]
\subfloat{\includegraphics[width=1\columnwidth]{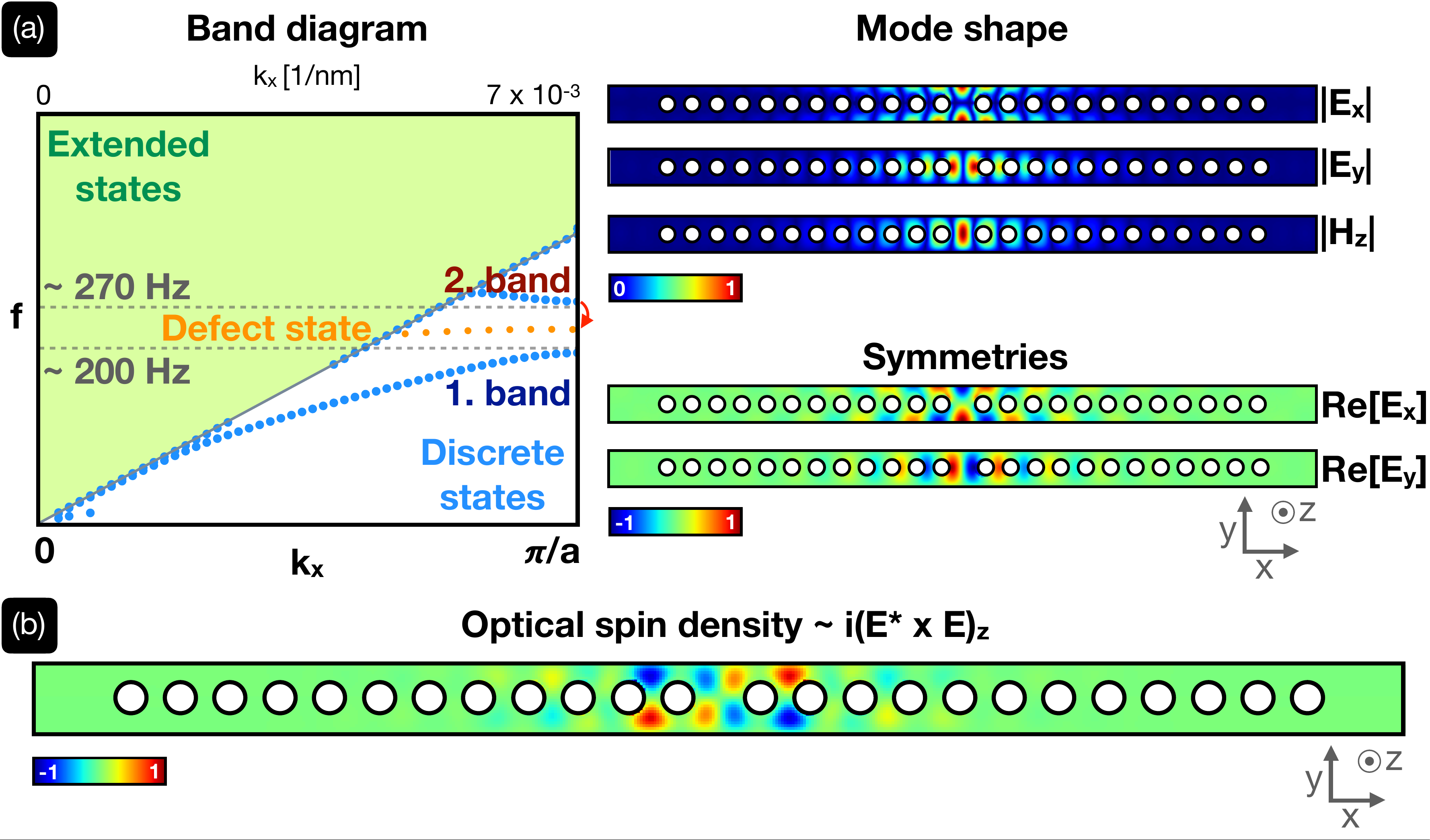}}\subfloat{\includegraphics[width=1\columnwidth]{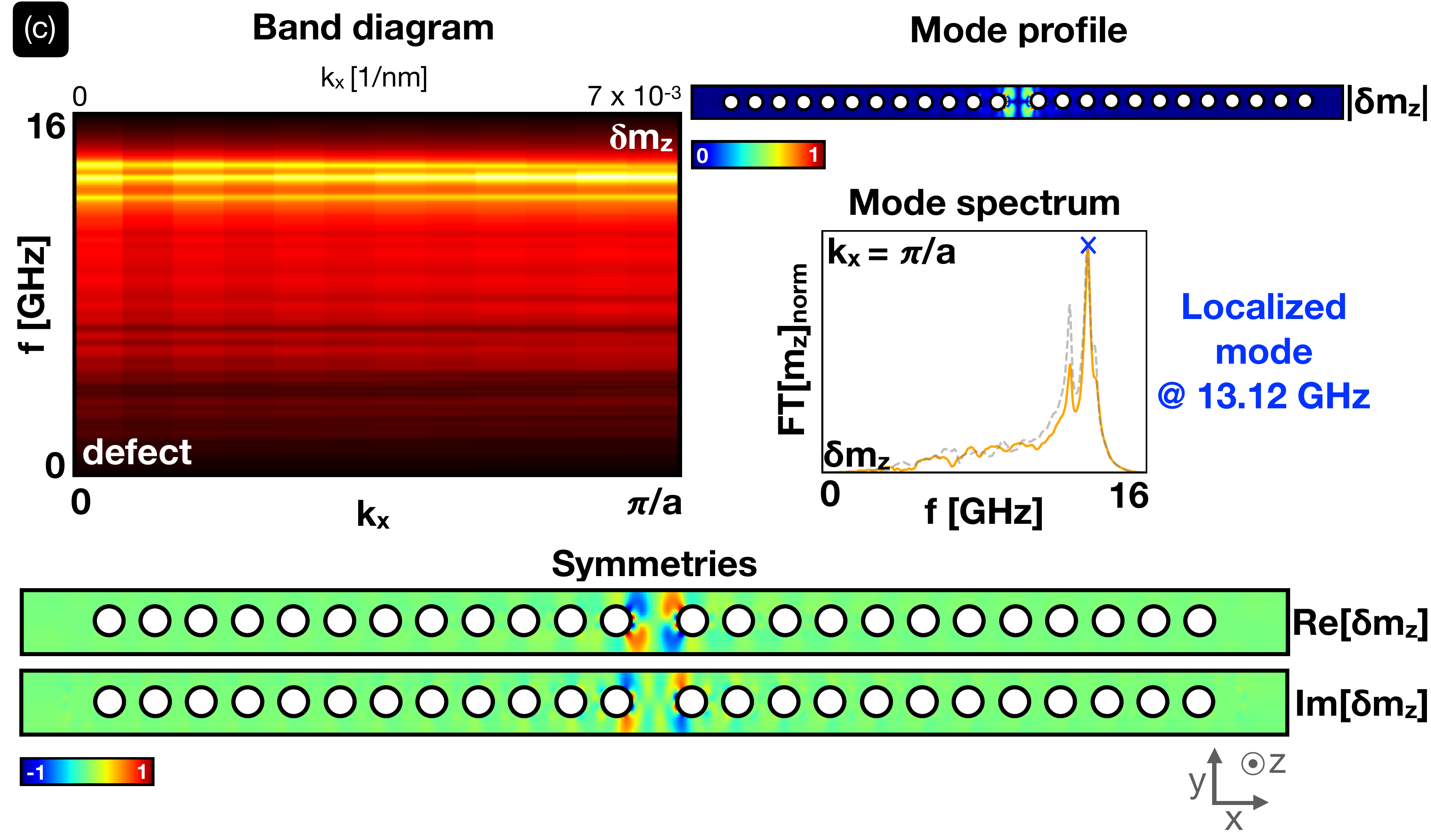}}
\caption{Optical (a and b) and magnetic modes (c): (a) Band diagram (obtained with MEEP) for TE-like modes within the irreducible BZ with a state that was pulled into the gap from the upper band-edge state by the insertion of a defect (note that the gap-state was not obtained by band diagram simulations). The bands in the green shaded area representing the light cone are leaky modes which couple with radiating states inside the light cone~\cite{MEEP_tutorial}. From the shape of the localized defect mode (obtained with Comsol) with a frequency of $\omega_{\text{opt}}= 2\pi \times \SI{246}{\tera\hertz}$ (middle layer in the $xy$-plane) we see that this mode is odd with respect to $x=0$ and $y=0$ (and even with respect to $z=0$). (b) Optical spin density (middle layer in the $xy$-plane) of the localized mode, fulfilling the same symmetries, see main text. (c) Band diagram of backward volume waves within the irreducible BZ showing magnetic modes with extended $\boldsymbol{k}$-values but preferring wave vectors at the edge of the BZ. The highest excited localized mode has a frequency of $\omega_{\text{mag}} = 2 \pi \times \SI{13.12}{\giga\hertz}$ and  is odd along the mirror symmetry planes for $x=0$ and $y=0$ (and additionally even with respect to the plane for $z=0$). The dashed line in the middle inset shows the mode spectrum in case of no defect.\\
Note: all mode shape plots are normalized to their corresponding maximum value.}
 \label{figure_4}
\end{figure*}

For evaluating the optical modes we used the two finite element tools MEEP~\cite{MEEP} and Comsol Multiphysics~\cite{comsol} (see appendix B).
The simulated band structure for TE-like modes in the considered photonic crystal shows a broad band gap in the infra-red frequency range with a nicely pulled defect band (see Fig.~\ref{figure_4}(a)) which is extended in frequency space resulting from the confinement in real space.
The defect mode in the gap at the edge of the BZ has a frequency of $\omega_{\text{opt}} \approx 2\pi \times \SI{246}{\tera\hertz}$ (obtained by Comsol, $\SI{205}{\tera \hertz}$/ $\lambda \approx \SI{1550}{\nano\metre}$ according to MEEP, note that the difference is due to a reduction of the simulation geometry to 2D in order to save simulation time) with a damping factor of $\kappa_{\text{opt}} \approx 2\pi \times \SI{0.1}{\tera\hertz}$ which gives a linewidth (FWHM) of $\gamma_{\text{opt}} = 2 \, \cdot \, \kappa_{\text{opt}} \approx 2\pi \times \SI{0.2}{\tera\hertz}$.
Using the values obtained by Comsol this gives an optical quality factor of $\mathcal{Q} = \omega_{\text{opt}}/(4\pi \kappa_{\text{opt}} ) = 1250$, which is in the expected range for this kind of geometry \cite{optomagnonic_crystal_idea} (note that MEEP gives a roughly three times larger value due to the 2D simulation which effectively resembles a simulated system of infinite height). The obtained quality factor is however small compared to 1D crystals made of silicon with a smooth defect, where quality factors in the order of $10^4 - 10^6$ can be achieved~\cite{opt_mech_crys_1D_7, opt_mech_crys_1D_5, opt_mech_crys_1D_3}. 

The corresponding mode shape in real space is shown in Fig.~\ref{figure_4}(a). We see that the mode is nicely localized at the defect.
Furthermore the $E_x$ component is even (odd) as a function of $x$ ($y$), whereas the $E_y$ is even as a function of both $x$ and $y$ fulfilling the symmetry requirements for a TE-like mode given in Eqs.~\eqref{TE-like_mode}.
 Due to this symmetry, we can disregard the $E_z$ component here since $E_z \approx 0$.  For the Faraday component of the optomagnonic coupling, the relevant quantity is the electric component of the optical spin density, $\boldsymbol{S}_{\rm opt} \propto \boldsymbol{E}^* \times \boldsymbol{E}$ [see Eq.~\eqref{optomagnonic_coupling}].
$\boldsymbol{S}_{\rm opt}$ points mostly along $z$-direction, is odd as a function of $x$ and $y$, and is even along $z$ (see Fig.~\ref{figure_4}(b) and \ref{figure_6}(b)).

\section{Magnonic crystal} 
\label{magnonic_crystal_sec}
As photonic crystals control the flow of light, magnonic crystals can be used to manipulate the spin wave dynamics in magnetic materials.
In general a magnonic crystal is made of a magnetic material with a periodic distribution of material parameters. 
Examples include the modulation of  the saturation magnetization or the magnetocrystalline anisotropy, a periodic distribution of different materials, or the modulation by external parameters, such as an applied magnetic field~\cite{magnonic_crystals_magnonics,  magnonic_crystals_review_and_prospects, magnonic_crystals_building_blocks_of_magnonics}.
Historically, magnonic crystals precede photonic crystals~\cite{magnonic_crystal_history_1, magnonic_crystal_history_2}. 
Unlike in photonic or phononic crystals, the band structure in magnonic crystals depends not only on the periodicity of the crystal but also on the spatial arrangement of the ground state magnetization, resulting in an additional degree of freedom.
Hence the band structure depends on the applied external magnetic field, the relative direction of the wave vector, the shape of the magnet, and the magnetocrystalline anisotropy of the material~\cite{magnonic_crystals_magnonics,  magnonic_crystals_review_and_prospects, magnonic_crystals_building_blocks_of_magnonics}. 
In this section we study the properties of the crystal presented in Sec.~\ref{photonic_crystal_sec}  (see Fig.~\ref{figure_3}(a)), as a magnonic crystal.

\begin{figure}
\begin{centering}
\includegraphics[width=1\columnwidth]{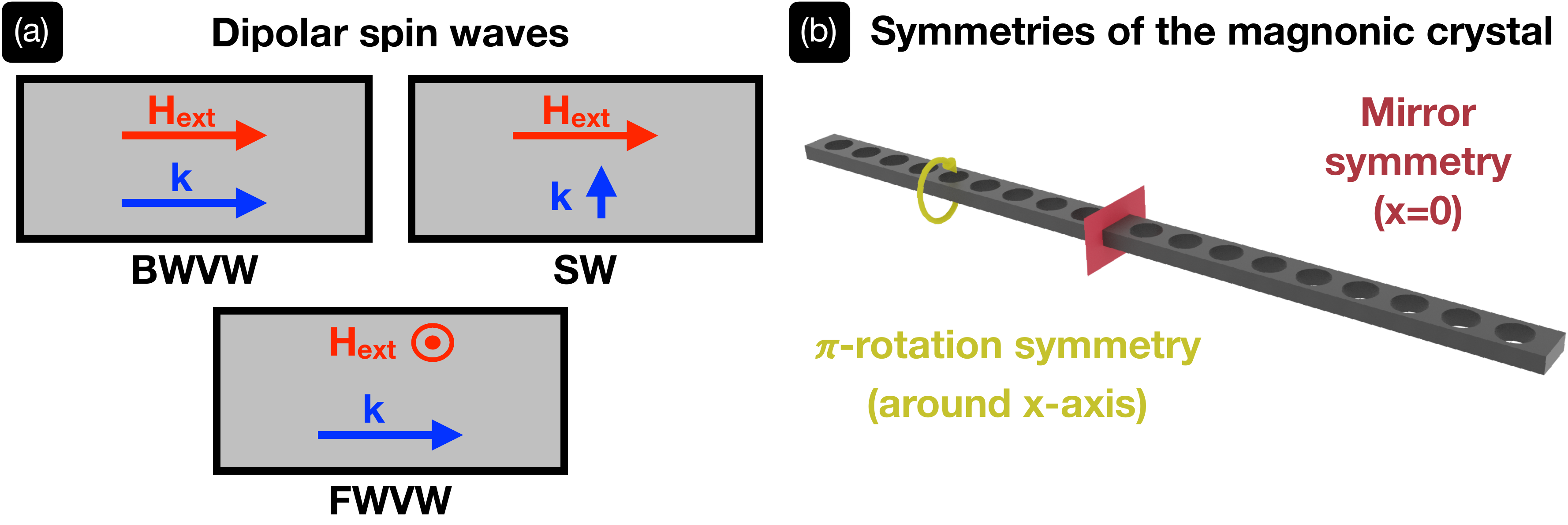}
\par\end{centering}
\caption{Dipolar spin wave types and symmetries in the magnonic crystal: (a) Dipolar spin waves can be divided into three types: backward volume waves (BWVW) with their wave vector parallel to the external field which both lie in the plane of the structure ($\boldsymbol{k} \, \| \, \boldsymbol{m}_{0} \, \| \, \boldsymbol{H}_{\textbf{ext}}$). Forward volume waves (FWVW) with their wave vector in plane and perpendicular to the external field which lies normal to the structure's plane ($\boldsymbol{k} \, \bot \, \boldsymbol{m}_{0} \, \| \, \boldsymbol{H}_{\textbf{ext}}$). Surface waves are also forward volume waves but they have their wave vector in plane and perpendicular to the external field which also lies in plane of the structure. ($\boldsymbol{k} \, \bot \, \boldsymbol{m}_{0} \, \| \, \boldsymbol{H}_{\textbf{ext}}$). (b) Symmetries of the investigated  1D magnonic crystal shown in Fig.~\ref{figure_1}(a). Since the external magnetic field breaks two mirror symmetry planes only the mirror symmetry plane normal to the saturation direction remains. Additionally a $\pi$-rotation symmetry around the saturation axis is present.}
 \label{figure_5}
\end{figure}

In the following we consider magnetic excitations which are non homogeneous in space, and we focus only on systems in the presence of an external magnetic field saturating the magnetization in a chosen direction. 
In this case spin waves can be divided into three classes: if all spins precess uniformly in phase, the mode is homogeneous and denominated the Kittel mode. 
If the dispersion is dominated by dipolar interactions (which is usually the case for wavelengths above $\SI{100}{\nano\meter}$) the excitations are called dipolar spin waves. 
For wavelengths below $\SI{100}{\nano\meter}$ the exchange interaction dominates instead, giving rise to exchange spin waves. 
The frequencies of the dipolar spin waves lie typically in the GHz-regime, whereas the exchange spin waves have frequencies in the THz-regime.
Since the  size of the structure considered in this work is in the micrometre range, we will focus on dipolar spin waves. For this case, the modes can be classified further by their propagation direction with respect to the magnetization. For an in-plane magnetic field,
modes with a frequency higher than the frequency of the uniform precession tend to localize at the surface and have a wave vector pointing perpendicular to the static magnetization $\boldsymbol{M}_{0}$ and thus the external field,  $\boldsymbol{k} \, \bot \, \boldsymbol{M}_{0} \, \| \, \boldsymbol{H}_{\textbf{ext}}$ (see Fig.~\ref{figure_5}(a)).
These modes are called surface or Damon-Eshbach modes~\cite{damon_eshbach, Serga_2010}.
If the wave vector is parallel to the external field such that $\boldsymbol{k} \, \| \, \boldsymbol{M}_{0} \, \| \, \boldsymbol{H}_{\textbf{ext}}$ holds, the waves are called backward volume waves and their frequency is smaller than the frequency of the Kittel mode (see Fig.~\ref{figure_5}(a)).
Finally, if the external field and the magnetization are normal to the crystal's plane and the wave vector lies in plane $\boldsymbol{k} \, \bot \, \boldsymbol{M}_{0} \, \| \, \boldsymbol{H}_{\textbf{ext}}$ these waves are called forward volume waves (see Fig.~\ref{figure_5}(a))~\cite{magnonic_crystals_building_blocks_of_magnonics, simulation_bands_1}. 
In the following we restrict the discussion to external fields which are applied in the plane of the crystal.

Similar to light modes in photonic crystals, magnon modes can also be localized within a certain region in the magnonic crystal. 
It is well known that the two dimensional periodic modification of a continuous film, for example by the  insertion of holes (denominated antidot arrays) can drastically change the behavior of the spin waves~\cite{magnonic_crystals_localization_1, magnonic_crystals_localization_2}.
In this case the modes have either a localized or extended character.
The localized mode is a consequence of non-uniform demagnetization fields created by the antidots. These fields change abruptly at the edges of the antidots and act as  potential wells for the spin waves~\cite{magnonic_crystals_building_blocks_of_magnonics}.
Thus, the above designed crystal, which localizes the optical mode by the insertion of a defect, is also a good candidate for acting as a magnonic crystal localizing magnetic modes via the holes.
Although the geometry of the crystal is optimized for the optics, it should be able to host and localize magnetic modes due to its shape and material (YIG).
Therefore we do not change the crystal further and use this structure as a proof of principle. This implies that we expect considerable room of improvement with respect to the optomagnonic coupling rates obtained in this structure. YIG is a good choice for magnonics since it has the lowest spin wave damping when compared to other materials commonly used~\cite{magnonic_crystals_review_and_prospects}. It is however difficult to pattern at the microscale, but recent advances in fabrication show great promise in this respect \cite{yig_fab_1,yig_fab_2}.

For concreteness, in the following we proceed to design the Faraday part of the optomagnonic coupling $G^{\text{F}}$, see Eq.~\eqref{optomagnonic_coupling}. Since $G^{\text{F}}$ is proportional to the overlap integral between the optical spin density and the magnon mode, we search for a magnon mode with the same symmetries as the optical spin density, in order to get the highest possible overlap. 
Like in the optical case, the magnonic crystal has three mirror symmetry planes ($z=0,~y=0,~x=0$).
However, the external applied magnetic field saturating the magnetization breaks two of these symmetries and thus only the mirror symmetry w.r.t. the plane perpendicular to the external field remains (see Fig.~\ref{figure_5}(b)). 
Note that the magnetization is a pseudo vector and its components perpendicular to the mirror does not change. Thus, the mirror operation is inverted from Eq.~(\ref{symmetry_defs}),
\begin{equation}
\hat{\sigma}^{\text{M}}_x \, \delta \boldsymbol{m}(\boldsymbol{r}) = \left( \begin{array}{c}
\delta m_x (-x,y,z) \\ 
-\delta m_y (-x,y,z) \\ 
-\delta m_z (-x,y,z)
\end{array}  \right) = \delta \boldsymbol{m}(\boldsymbol{r}).
\label{magnetics_x_symmetry}
\end{equation}
Since the optical spin density pointing along $\hat{z}$ is odd as a function of $x$, we require $\delta m_z$ to be odd as well and consequently $\delta\boldsymbol{m}$ to be even under $\hat{\sigma}^{\text{M}}_x$.
Additionally, a $\pi$-rotation around the $\hat{x}$-axis symmetry remains unbroken
\begin{equation}
\hat{\cal R}^{\pi}_x \, \delta \boldsymbol{m}(\boldsymbol{r}) = \left( \begin{array}{c}
\delta m_x (x,-y,-z) \\ 
-\delta m_y (x,-y,-z) \\ 
-\delta m_z (x,-y,-z)
\end{array}  \right) = \delta \boldsymbol{m}(\boldsymbol{r}).
\label{magnetics_x_symmetry}
\end{equation}
Invoking again the symmetries of the optical spin density (odd as a function of $y$ and even with $z$) we consider modes with even rotational symmetry.
We note that due to the different symmetries respected by the photon and magnon modes, we choose the symmetries of the modes in such a way that they preferably match in the $xy$-plane, which is the most relevant dimension for thin structures. In this case, the symmetries of the optical and the magnetic mode along the height do not necessarily match. For thin films however they do, see Fig.~\ref{figure_6}(b).

Since spin waves are excited by an external magnetic pulse which controls the direction of the wave vector $\boldsymbol{k}$, the pulse also breaks the mirror symmetries of the crystal.
Therefore we focus on a setup which conserves the relevant mirror symmetry, and only excite backward volume waves where the external saturation field and the wave vector of the mode are parallel and lie in the plane of the crystal.
We note that this configuration is also the most favourable one from an experimental point of view, and additionally the configuration most likely used in magnonic devices~\cite{simulation_bands_1}.

We evaluated the magnetization dynamics numerically by means of the finite difference tool MuMax3~\cite{mumax} which solves for the Landau-Lifshitz-Gilbert equation of motion for the local magnetization vector~(see appendix C).
In order to excite magnon modes with the desired symmetry, we use a 2D antisymmetric sinc-pulse which should moreover avoid spurious effects in the spectrum~\cite{simulation_bands_2} 
\begin{equation}
\begin{split}
\boldsymbol{H}_{\text{pulse}} = &\text{H}_{\text{pulse}} \frac{\sin^2(\omega_{\text{c}} t)}{\omega_{\text{c}} t} \frac{\sin^2(k_{\text{c}} x)}{k_{\text{c}} x}  \frac{\sin^2(k_{\text{c}} y)}{k_{\text{c}} y}  \boldsymbol{e}_y,
\end{split}
\end{equation}
pointing along the $\hat{y}$-direction in order to excite backward volume waves~\cite{simulation_bands_1}.
The cut-off frequency was chosen to be $\omega_{\text{c}} = 2\pi \times  \SI{16}{\giga\hertz}$ and the cut-off wave vector to be $k_{\text{c}} = \pi/a$ in order to concentrate all the excitation energy in the first BZ. 
Since this pulse is centered in the middle of the crystal, we only excite modes around the crystal's center.
The external saturation field was set to $\text{H}_{\text{ext}} = \SI{400}{\milli\tesla}$ (found by hysteresis) and the pulse field to $\text{H}_{\text{pulse}} = \SI{0.4}{\milli\tesla}$.
We note that the pulse strength should be a small perturbation of the saturation field in order to minimize non linear effects. 
We used the material parameters for YIG, $M_{\text{s}} = \SI{140}{\kilo\ampere/\metre}$ (saturation magnetization), $A_{\text{ex}} = \SI{2}{\pico\joule/\metre}$ (exchange constant), $K_{\text{c}1} = \SI{-610}{\joule/\metre^3}$ (anisotropy constant) with the anisotropy axis along $\hat{z}$~\cite{yig_parameters}.
In order to accelerate the simulations, we used an increased Gilbert damping parameter  $ \gamma = 0.008$ (compare to $\gamma\approx 10^{-5}-10^{-4}$ for YIG)~\cite{saga_of_yig, yig_damping}.

In the following considerations we focus only on the $\delta m_z$ component of the magnetization dynamics, since the optical spin density of a TE-like mode mostly points into the $\hat{z}$-direction, rendering $\delta m_x$ and  $\delta m_y$ irrelevant for $G^{\text{F}}$ (see Eq.~\eqref{optomagnonic_coupling}). We find that the optical defect also acts as confinement of the magnetic mode, resulting in the defect like dispersion relation presented in Fig.~\ref{figure_4}(c). 
The obtained band structure shows modes around the edge of the BZ with extended wave vector character, implying that the modes are highly localized in space.
The frequency of the highest excited localized mode at the BZ edge is $\omega_{\text{mag}} =2\pi \times \SI{13.12}{\giga\hertz}$ with an estimated linewidth (FWHM) of $\Gamma_{\text{mag}} = \gamma \, \omega_{\text{mag}} =  2\pi \times \SI{131.2}{\kilo\hertz}$ where we used the real Gilbert damping of YIG $\gamma = 10^{-5}$.
Note that the simulated linewidth shown in Fig.~\ref{figure_4}c is larger due to the different choice of the Gilbert damping in order to speed up the simulation.

As we see from its mode shape, this mode is nicely localized at the holes attached to the defect and is odd with respect to $x=0$ and $y=0$, and hence has the same symmetry as the optical spin density as we aimed for (see Fig.~\ref{figure_4}c).

\section{Optomagnonic crystal} \label{optomagnonic_crystal_sec}

As shown above, the crystal in Fig.~\ref{figure_1}a can host both optical and magnetic modes and therefore can be considered an \emph{optomagnonic crystal}. 
In this section, we evaluate the optomagnonic coupling $G^{\text{F}}$ given in Eq.~(\ref{optomagnonic_coupling}) ($G^{\text{C}}$ is briefly discussed at the end of the section) for the modes found in Secs.~\ref{photonic_crystal_sec} and~\ref{magnonic_crystal_sec} shown in Fig.~\ref{figure_4}.
\begin{figure}[t!]
\includegraphics[width=1\columnwidth]{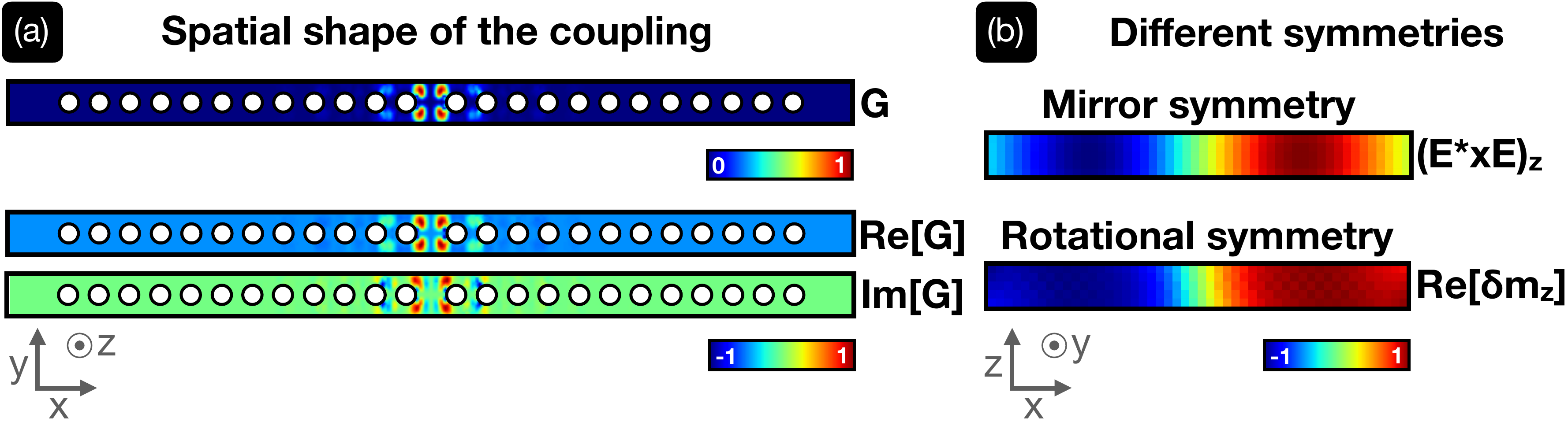}
\caption{Spatial shape of the coupling and different symmetries of the optical and the magnetic mode: (a) Spatial shape of the coupling similar to the magnon mode shape. (b) Different symmetries along the crystal's height of the optical spin density and the magnon mode. Due to the external magnetic field the mirror symmetry along the height is broken and only a $\pi$-rotation symmetry remains, resulting in different mode shapes along the height of the crystal. For thin films this difference is rather small.\\
Note: all mode shape plots are normalized to their corresponding maximum value.}
 \label{figure_6}
\end{figure}

\begin{figure}[t!]
\includegraphics[width=1\columnwidth]{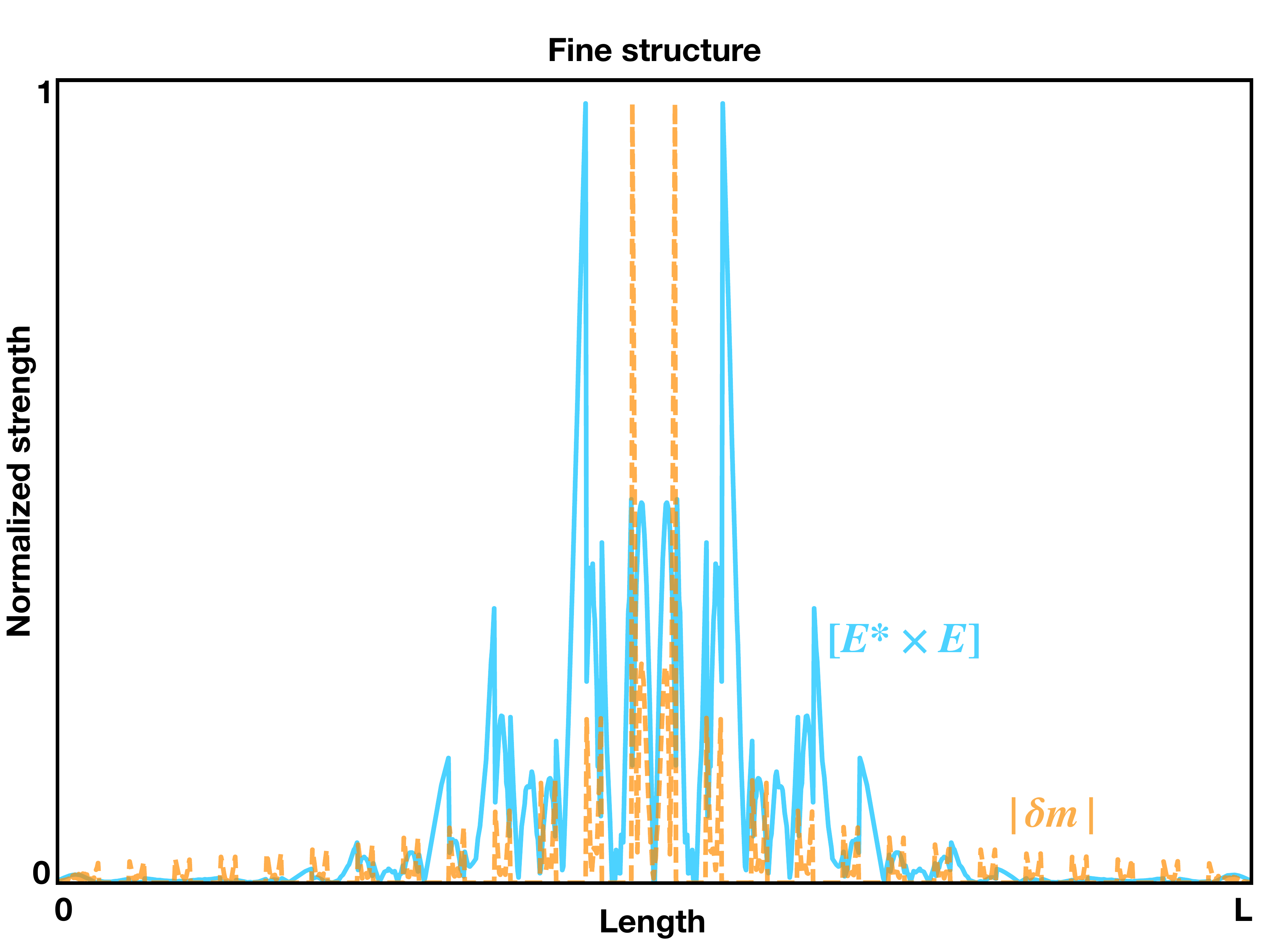}
\caption{Fine structure of the optical spin density and the magnon mode along the length of the crystal for a fixed height and width. }
 \label{figure_10}
\end{figure}

Numerically evaluating Eq.~\eqref{optomagnonic_coupling} gives a Faraday contribution to the optomagnonic coupling per magnon and per photon of $|G_{\text{num}}^{\text{F}}| = 2 \pi \times \SI{0.5}{\kilo\hertz}$ (spatial shape of the coupling see Fig.~\ref{figure_6}a).
In order to gauge this value we want to compare it to the analytical estimate derived in~\cite{optomagnonics_review}.
In the optimal case, the magnetic mode volume and the optical mode volume coincide, $V_{\text{mag}} \approx V_{\text{opt}}$. 
In this case, we estimate the coupling as
\begin{equation}
|G_{\text{optimal}}^{\text{F}}| = \frac{\theta_{\text{F}} \lambda_{n}}{2 \pi} \, \omega_{\text{opt}} \, \sqrt{\frac{g \mu_{\text{B}}}{M_{\text{s}}}} \, \frac{1}{\sqrt{\text{V}_{\text{mag}}}},
\label{optomagnonic_coupling_estimate}
\end{equation}
which evaluates to $|G_{\text{optimal}}^{\text{F}}| = 2\pi \times \SI{0.6}{\mega\hertz}$ using the material parameters of YIG ($(\theta_{\text{F}} \lambda_{n})/(2 \pi) = 4 \cdot 10^{-5},~ M_{\text{s}} = \SI{140}{\kilo\ampere/\metre}$) and the optical frequency found in Sec.~\ref{photonic_crystal_sec}, $\omega_{\text{opt}} = 2\pi \times \SI{249}{\tera \hertz}$. The magnetic mode volume is defined as the one where the magnon intensity is above a certain threshold, giving $V_{\text{mag}} = 2.8\cdot 10^{-2} \SI{}{\micro\meter^3}$ (see appendix D).
The coupling is bounded by the magnon mode volume, since in the investigated structure it is smaller than the optical mode volume (see Fig.~\ref{figure_4}).
In order to take the mismatch in the mode volume into account, we introduce the following overlap measure which is also known as filling factor
\begin{equation}
\mathcal{O} = \frac{V_{\text{overlap}}}{V_{\text{opt}}},
\label{overlap_measure}
\end{equation}
where $V_{\text{overlap}}$ represents the volume where the magnon and photon modes overlap. The volumes are estimated similar to the case of magnons to be $V_{\text{overlap}} = 9.7 \cdot 10^{-3}\SI{}{\micro\meter^3}$ and $V_{\text{opt}} = \SI{0.7}{\micro\meter^3}$ (see appendix D). 
Note that for the optical volume it was taken into account that the mode leaks out of YIG into the $\text{Si}_3\text{N}_4$ layer and air, which is not shown in Fig.~\ref{figure_4} (a)+(b).
Thus the overlap measure evaluates to $\mathcal{O}=0.01$, shrinking the optimal coupling to $\mathcal{O} \cdot |G_{\text{optimal}}^{\text{F}}| \approx 2 \pi \times \SI{6}{\kilo\hertz}$.
Hence, even though the optomagnonic crystal localizes both modes in the same region, the overlap measure is rather small due to the much larger optical mode volume (see Fig.~\ref{figure_4} and Fig~\ref{figure_6}(a)), which is detrimental for the coupling strength. 
Furthermore by looking at the fine structure of the optical spin density and the magnon mode we see that the amplitude peaks of both do not coincide (see Fig.~\ref{figure_10}): the magnonic peaks are localized nearer to the center than the optical ones. This results in a smaller overlap volume which would be $~V_{\text{mag}}$ if the peaks of the modes would be at the same position.

Since the coupling also strongly depends on the relative direction between the vectors of the modes, we additionally introduce a `directionality' measure
\begin{equation}
\mathcal{D} =   \frac{\int \text{d} \boldsymbol{r} ~ \delta \boldsymbol{m} (\boldsymbol{r}) \cdot [\boldsymbol{E}^* (\boldsymbol{r}) \, \times \, \boldsymbol{E} (\boldsymbol{r})]}{\int \text{d} \boldsymbol{r}~|\delta \boldsymbol{m} (\boldsymbol{r})| \, |\boldsymbol{E}^* (\boldsymbol{r})\times \boldsymbol{E} (\boldsymbol{r})|}
\label{direction_measure}
\end{equation}
evaluating to $\mathcal{D} = 51 \%$ using the numerical results presented above.
As we see, although the symmetries of the optical spin density and the magnon mode match, the vectors of the modes do not perfectly align in the defect area (see Figs.~\ref{figure_4}).
Taking also this sub-optimal alignment into account the coupling estimate reduces to $|G_{\text{expected}}^F| = \mathcal{O} \cdot \mathcal{D} \cdot |G_{\text{optimal}}^{\text{F}}| = 2 \pi \times \SI{3}{\kilo\hertz}$ which coincides well with the numerically obtained value.
We conclude that the coupling in the investigated structure is mostly affected by the large difference between the optical and the magnetic mode volumes, shrinking the coupling value by two orders of magnitude. We remind the reader that the obtained values are for a proof of principle structure which has been only partially optimized, since we started from a fixed photonic crystal structure. In the next section we discuss a possible optimization from the magnonics side.

We now proceed to briefly discuss the Cotton-Moutton effect for the results found in Secs.~\ref{photonic_crystal_sec} and~\ref{magnonic_crystal_sec}. For YIG, the Cotton-Moutton coefficient $(\theta_{\text{C}} \lambda_{n})/(2\pi) = - 2 \times 10^{-5}$~\cite{faraday_2} is of the same order of magnitude as the corresponding Faraday coefficient, determined by $\theta_{\text{F}}$. Since in the Voigt configuration both effects are of leading order in the magnetization fluctuations (see Eqs.~\eqref{optomagnonic_coupling} and~ \eqref{CM_coupling}), it is important to take its contribution into account. Moreover, since the coefficients $G
^{\text{F}}$ and $G^{\text{C}}$ are complex, it is difficult to estimate a priori the total coupling $|G
^{\text{F}}+G^{\text{C}}|$, due to the unknown possible interference effects. Numerically evaluating Eq.~\eqref{CM_coupling} gives an interaction value of $|G^{\text{C}}_{\text{num}}| = 2\pi \times \SI{1.6}{\kilo \hertz}$. This large value can be explained by the symmetry of the integrand which reduces to $m_0^x [ E_x E_y^* \delta m_y + E^*_x E_y \delta m_y ]$ due to the backward volume wave setup and the TE-like character of the optical mode.
This integrand is fully even since $E_x E_y^*$ has the same symmetry as $\delta m_y$.
The full optomagnonic coupling 
\begin{equation}
\left|G_{\text{num}}\right| = \left|G^{\text{F}}_{\text{num}} + G^{\text{C}}_{\text{num}}\right|
\end{equation}
is found to be $G_{\text{num}} = 2\pi \times \SI{1.3}{\kilo\hertz}$.

Compared to the optomechanical coupling in similar 1D crystals, where coupling values (per photon and phonon) up to $2 \pi \times \SI{950}{\kilo\hertz}$ can be obtained~\cite{opt_mech_crys_1D_8,opt_mech_crys_1D_7,opt_mech_crys_1D_6,opt_mech_crys_1D_5,opt_mech_crys_1D_4,opt_mech_crys_1D_3}, the optomagnonic coupling obtained here is still rather small. However, this is large compared to other optomagnonic systems. As we argued above, the coupling is limited by the imperfect spatial matching of magnons and photons with overlap ${\cal O} = 0.01$ while it is enhanced due to small volumes, $V_{\text{mag}} \sim \SI{0.01}{\micro\meter^3}$ and $V_{\text{opt}} \sim \SI{1}{\micro\meter^3}$. In the standard setups involving spheres \cite{optomagnonics_zhang,optomagnonics_osada,optomagnonics_haigh}, typically optical volumes are very large $\sim 10^5 \SI{}{\micro\meter^3}$ with low optomagnonic overlap $\sim 10^{-3}$, resulting in low couplings $\sim \SI{1}{\hertz}$. It was theoretically shown that $>75\%$ overlap in such systems is achievable \cite{optomagnonics_sharma_2} but the couplings would still be $\sim2\pi\times\SI{500}{\hertz}$. The miniaturization of an optical cavity to $\sim\SI{100}{\micro\meter^3}$ was demonstrated in ~\cite{james_small_cavity}, where the coupling is however still small, $2\pi\times\SI{50}{\hertz}$, in this case due to the large magnon volume involved.

An important prerequisite for applications in the quantum regime such as magnon cooling, wavelength conversion, and coherent state transfer based on optomagnonics is a high cooperativity. The cooperativity per photon and magnon is an important figure of merit which compares the strength of the coupling to the lifetime of the coupled modes, and is given by 
\begin{equation}
\mathcal{C}_0 = \frac{4 G_{\text{num}}^2}{\gamma_{\text{opt}} \Gamma_{\text{mag}}},
\end{equation}
where  $\gamma_{\text{opt}}$ is the optical linewidth (FWHM), and $\Gamma_{\text{mag}}$ is the magnonic linewidth (FWHM). 

To evaluate the theoretical cooperativity of the structure proposed in this manuscript, we use $\Gamma_{\text{mag}} = \gamma \omega_{\text{mag}}$ where $\gamma = 10^{-5}$ is the Gilbert constant and $\omega_{\text{mag}} = 2\pi \times \SI{13.12}{\giga \hertz}$. The optical linewidth is found from simulations to be $\gamma_{\text{opt}} = 2\pi \times  \SI{0.2}{\tera\hertz}$. 
\begin{figure}[t!]
\includegraphics[width=1\columnwidth]{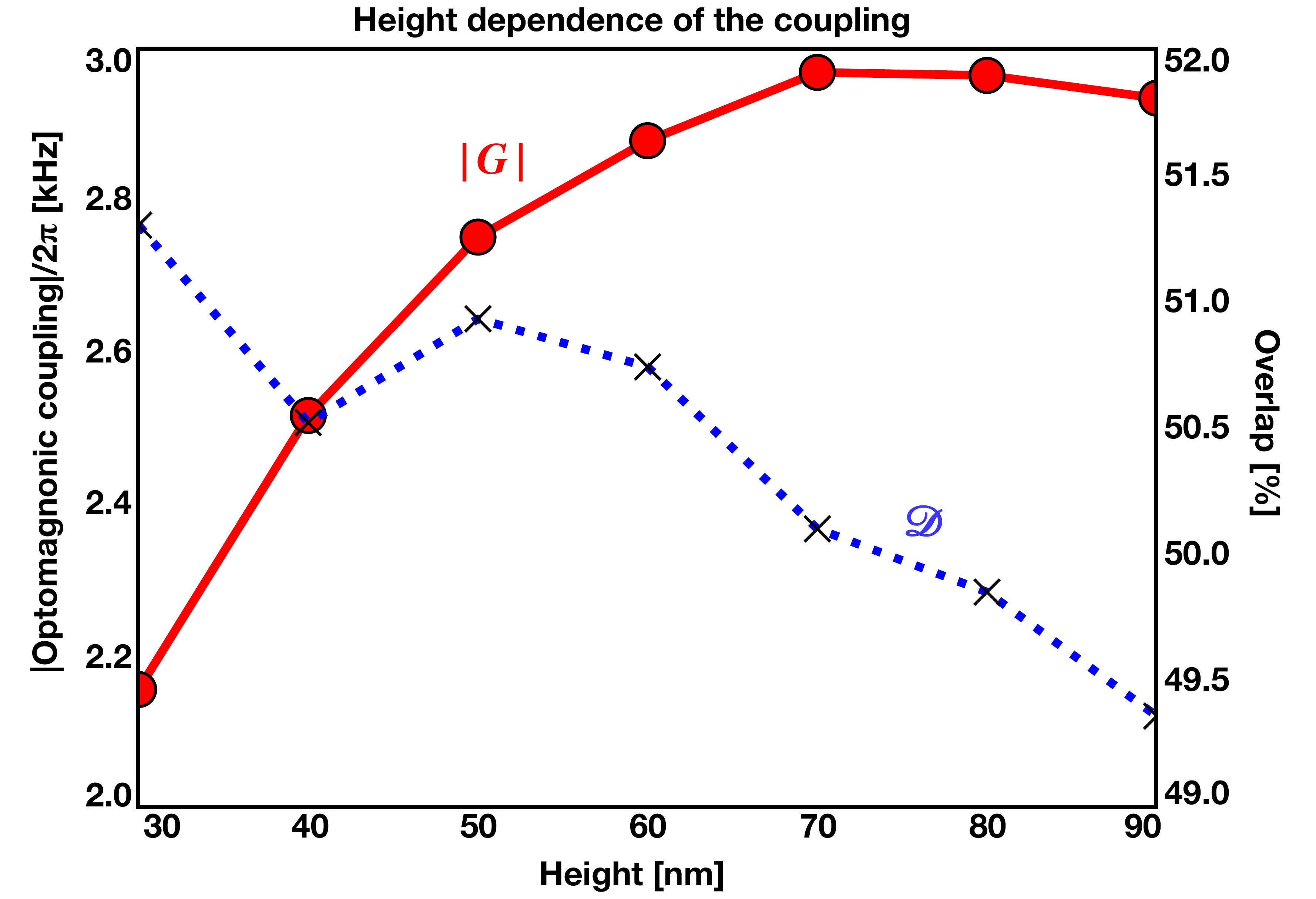}
\caption{Height dependence of the Faraday component of the optomagnonic coupling: The coupling shows a $\sqrt{V_{\text{mag}}}$ dependence since the optical mode volume in the YIG and the $\text{Si}_3\text{N}_4$ slab is constant. The decrease with larger height can be explained by the shrinking directionality measure (see Eq.~\ref{direction_measure}) between the optical and the magnetic mode.}
 \label{figure_9}
\end{figure}

Using the corresponding parameters the cooperativity per photon and magnon of the optomagnonic crystal is $\mathcal{C}^{\text{crystal}}_0 \sim 2.5 \cdot 10^{-10}$.
The single-particle cooperativity can be enhanced by the photon number in the cavity, $\mathcal{C} = n_{\text{ph}} \, \mathcal{C}_0$. Experimentally, there is a bound on the photon density that can be supported by the cavity without undesired effects due to heating, and it is empirically given by $ 5 \cdot 10^4$ photons per $\mu\mathrm{m}^3$~\cite{optomagnonic_new_4}. In our structure, considering the effective mode volume $V_{\text{opt}}$ this gives an enhanced cooperativity at maximum photon density of  $\mathcal{C}_{\text{crystal}} \sim 1 \cdot 10^{-5}$, which is two orders of magnitude larger than the current experimental state of the art~\cite{optomagnonic_new_4, james_small_cavity}.
 
Since our model does not account for fabrication imperfections, this number is expected to be lower in a physical implementation, indicating that optimization is needed. Results for similar 1D optomechanical crystals indicate that optimization can lead to larger cooperativity values (at maximum photon density), e.g. $\sim 10$~\cite{opt_mech_crys_1D_4}. The small cooperativity obtained in our structure is a combination of a reduced coupling due to mode-mismatch, plus the very modest quality factor of the optical mode in this simple geometry.
\begin{figure}[b!]
\includegraphics[width=1\columnwidth]{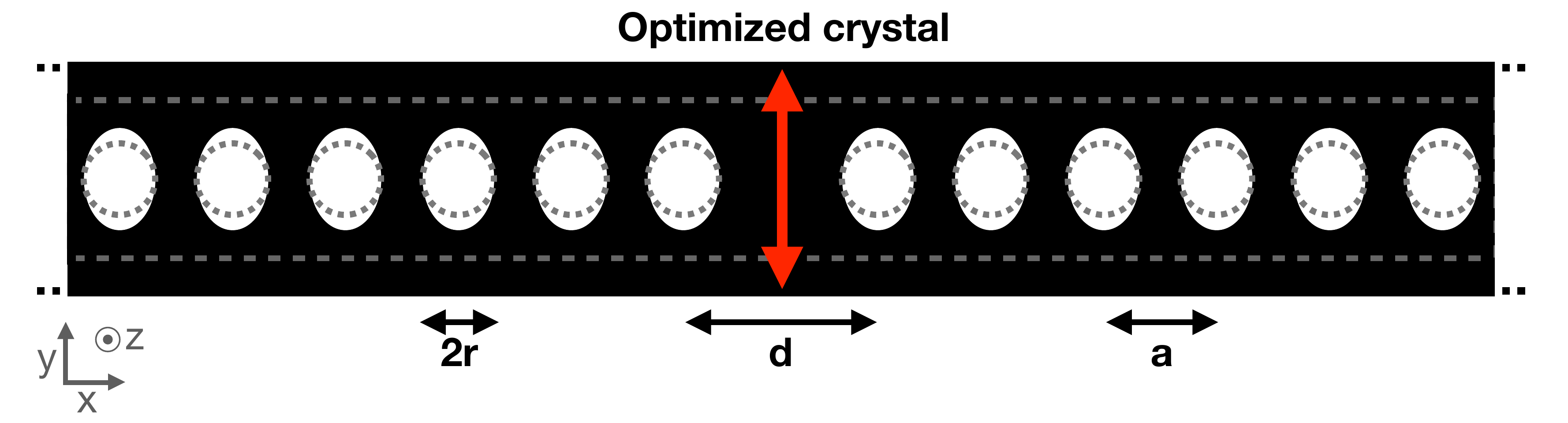}
\caption{ Optimization of the geometry: Through increasing the parameters along the width of the crystal we create more space for the modes without touching the optical optimization of the original crystal (dashed line). We note that we also increased the defect size, not shown here.}
 \label{figure_7}
\end{figure}

For boosting the coupling strength we investigate briefly in the following the influence of the optomagnonic crystal's height on the coupling, as proposed in Ref.~\cite{my_paper}.
Therefore we increase the height of the YIG layer from $\SI{30}{\nano\metre}$ to $\SI{90}{\nano\metre}$ without changing the other parameters of the geometry (including the $\text{Si}_3\text{N}_4$ layer in the optical simulations). 
As we see from the result (see Fig.~\ref{figure_9}) the coupling exhibits a $\sqrt{V_{\text{mag}}}$ dependence.
We find that the optical mode volume does not change substantially in the modified geometry, and therefore the observed behavior is consistent with the expected  $\sqrt{V_{\text{mag}}}/V_{\text{opt}}$ dependence for a constant optical mode volume. The slight decrease for larger heights can be explained by the shrinking of the  directionality measure $\mathcal{D}$, stemming from the difference in symmetries obeyed by the magnetization (rotational) and the electric field (mirror).
\begin{figure*}[t!]
\subfloat{\includegraphics[width=1\columnwidth]{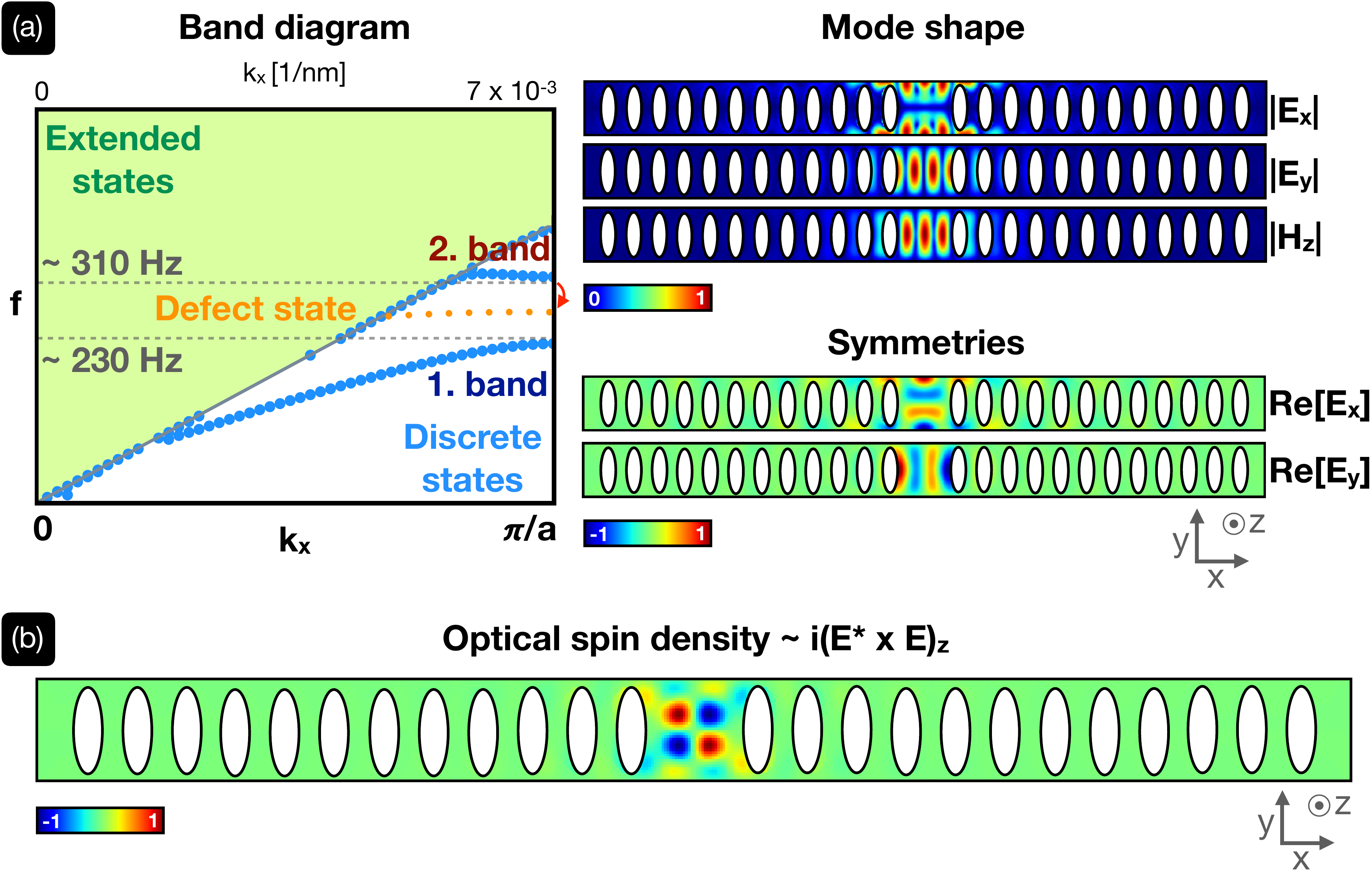}}\subfloat{\includegraphics[width=1\columnwidth]{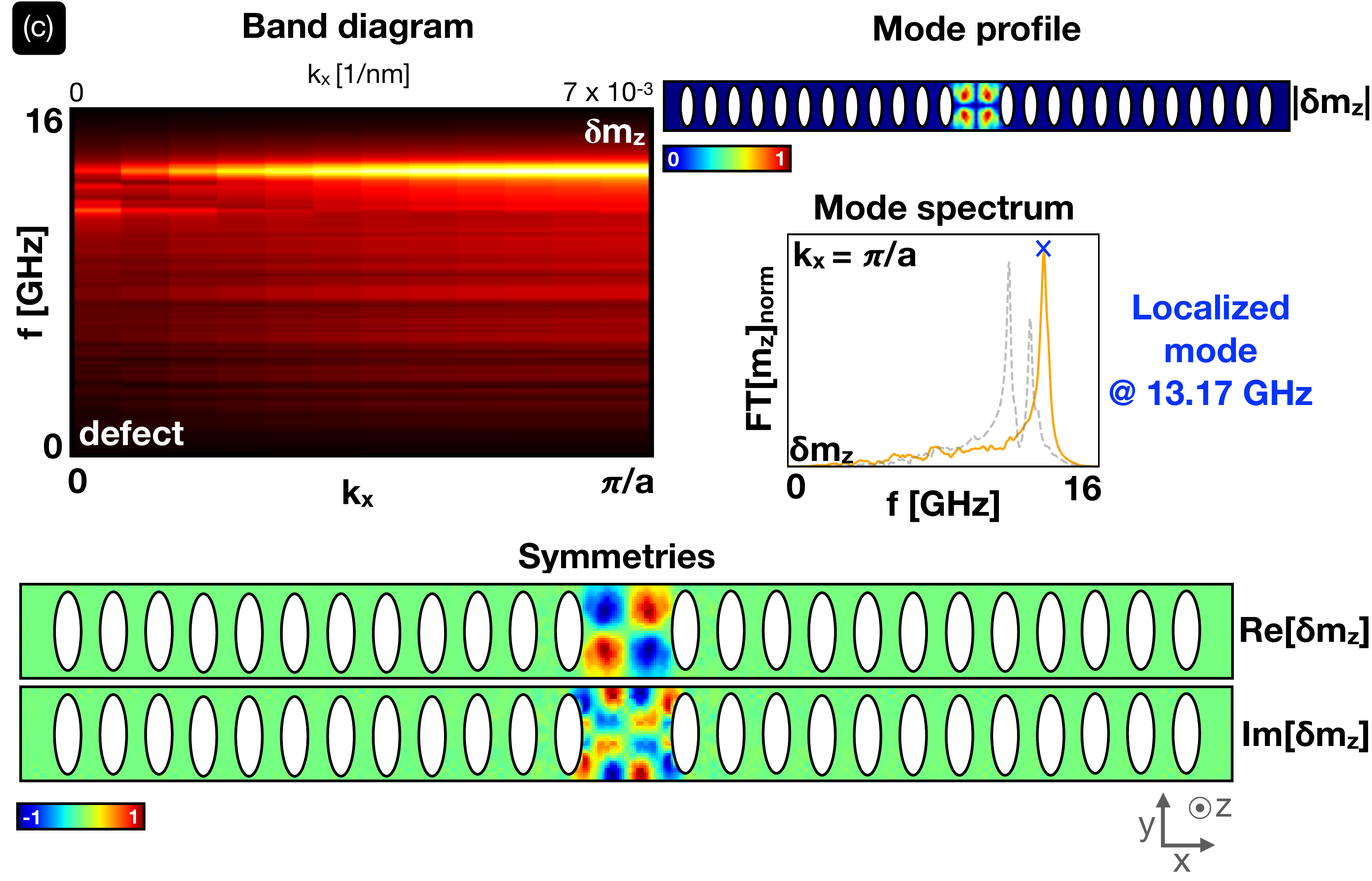}}
\caption{Optical (a and b) and magnetic modes (c) of the optimized crystal: (a) Band diagram for TE-like modes within the irreducible BZ with a defect mode in the photonic band gap which was pulled from the upper band-edge state into the gap by the insertion of a defect. From the mode shape of the localized mode with a frequency of $\omega_{\text{opt}}/2 \pi = \SI{279}{\tera\hertz}$ (middle layer in the $xy$-plane) we see that this mode is odd with respect to $x=0$ and $y=0$ (and even with respect to (z=0)). (b) Optical spin density of the localized mode (middle layer in the $xy$-plane) which is odd with respect to $x=0$ and $y=0$ (and even with respect to $z=0$). (c) Band diagram of backward volume waves within the irreducible BZ showing magnetic modes with extended $\boldsymbol{k}$-values but preferring wave vectors at the edge of the BZ. The highest excited localized mode has a frequency of $\omega_{\text{mag}} = 2 \pi \times \SI{13.17}{\giga\hertz}$ and  is odd along the mirror symmetry planes for $x=0$ and $y=0$ (and additionally even with respect to the plane for $z=0$). The dashed line in the middle inset shows the mode spectrum in case of no defect.\\
Note: all mode shape plots are normalized to their corresponding maximum value.}
 \label{figure_8}
\end{figure*}

\section{Optimization} \label{optimization_sec}
So far, we optimized the crystal in order to minimize optical losses for the given geometry. In this section we investigate how to optimize the geometry for magnonics. The optical optimization was achieved by fixing the hole radius and intra-hole distance, which are both along the length of the crystal. In the following we tune instead only the parameters along the width of the crystal ($\hat{y}$-direction), in order to perturb as little as possible the optical optimization. We found a promising structure by increasing the width of the crystal and considering elliptical holes, see Fig.~\ref{figure_7}. From a set of trials, we found that a width of $w = \SI{900}{\nano\meter}$ and a radius of the holes along the width of $r_w = \SI{380}{\nano\metre}$ give the highest coupling. An increased defect size of $d = \SI{1201.5}{\nano\meter}$ is also beneficial for decreasing the optical losses in this case, it nicely localizes the optical defect mode in the middle of the band gap and thus does not drastically change the localization behavior of the photonic crystal. 

For evaluating the photonic band structure and the optical modes we use the same procedure as described in Sec.~\ref{photonic_crystal_sec}. 
We obtain a similar band structure for TE-like modes and also a similar localized mode with a frequency of $\omega_{\text{opt}} = 2\pi \times  \SI{279}{\tera\hertz}$ (obtained by Comsol, $\SI{235}{\tera \hertz}$ according to MEEP) and a damping of $\kappa_{\text{opt}} = 2\pi \times  \SI{3}{\tera\hertz}$ which gives an optical linewidth (FWHM) of $\gamma_{\text{opt}} = 2\pi \times  \SI{6}{\tera\hertz}$ (see Fig.~\ref{figure_8}(a)).
Using the values obtained by Comsol this results in a reduced optical quality factor of $\mathcal{Q} = 93$ (note that MEEP gives a twice as large value). This rather low optical quality factor is a trade off for the magnetic optimization achieved by elliptical holes.
Moreover, the optical spin density compared to the original crystal is mostly localized within the defect which is advantageous for our purposes (see Fig.~\ref{figure_8}(b)). Similarly, for evaluating the magnon modes we used the parameters and procedures presented in Sec.~\ref{magnonic_crystal_sec}.
In the following we focus on the Faraday part of the optomagnonic coupling and therefore consider only the  $\delta m_z$ component of the magnon mode due to the structure of the optical spin density. The Cotton-Mouton term is discussed briefly at the end of the section.
The simulated band diagram for backward volume waves again shows extended magnon modes but in this case we obtain one broad band, most likely stemming from a fusion of several bands due to the larger width of the crystal (see Fig.~\ref{figure_8}(c)). 
The frequency of the highest excited localized mode is $\omega_{\text{mag}} = 2 \pi \times \SI{13.17}{\giga\hertz}$ with an estimated linewidth of $\gamma \omega_{\text{mag}} = 2 \pi \times \SI{131.7}{\kilo\hertz}$ where we used the Gilbert damping of YIG $\gamma = 10^{-5}$. As in the previous case,  the simulated linewidth is larger due to the larger Gilbert damping used in the simulations. 
As we see from its mode shape, this mode is nicely localized at the holes attached to the defect and has approximately the same shape and symmetry as the optical spin density (see Fig.~\ref{figure_8}(c)).

\begin{figure}[!]
\includegraphics[width=1\columnwidth]{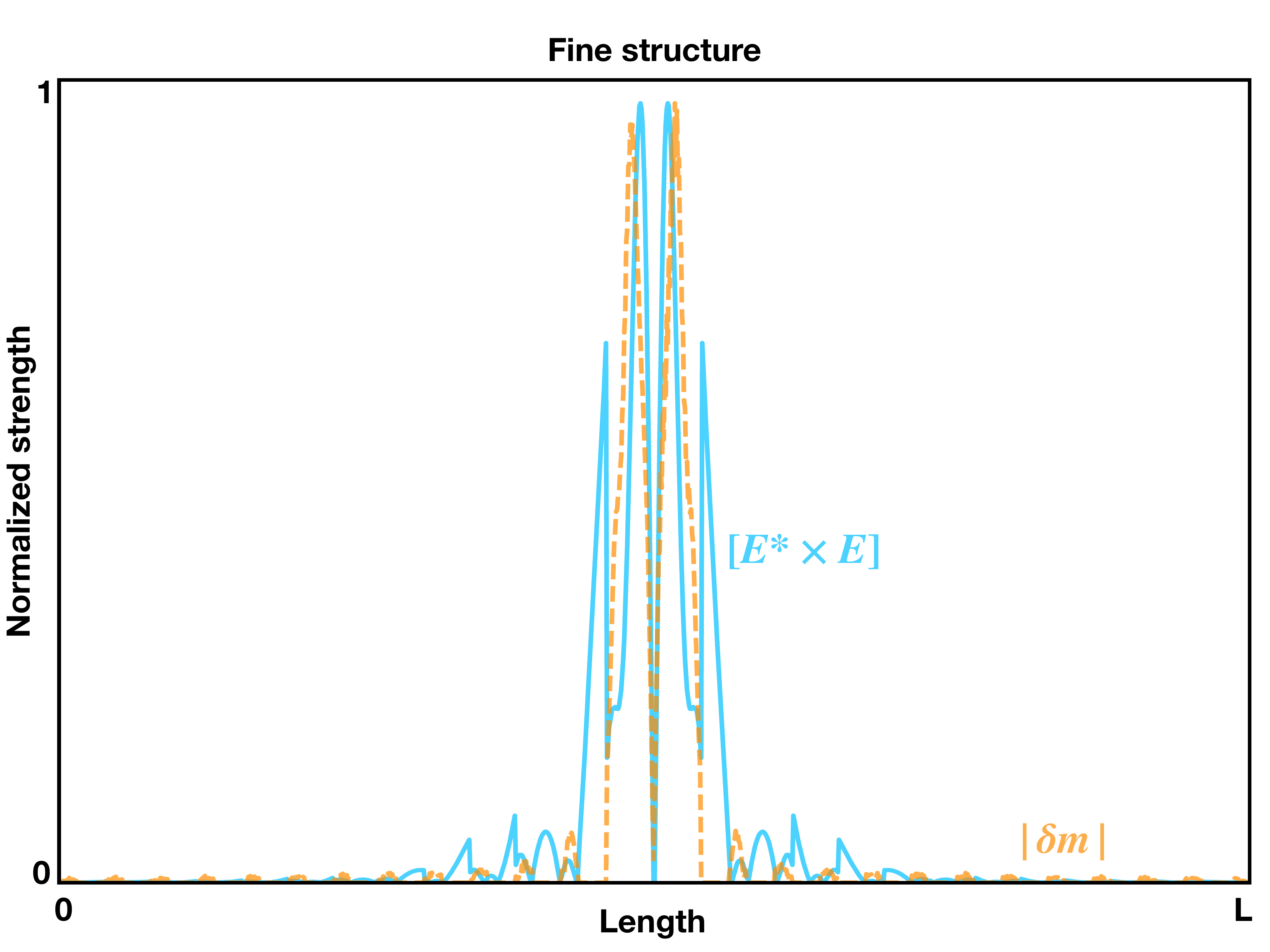}
\caption{Fine structure of the optical spin density and the magnon mode along the length of the optimized crystal for a fixed height and width. }
 \label{figure_11}
\end{figure}

Using the results discussed above, the Faraday component of the optomagnonic coupling of Eq.~\ref{optomagnonic_coupling} for the optimized crystal evaluates to $|G^{\text{F}}_{\text{num}}| = 2\pi \times \SI{2.9}{\kilo\hertz}$. Therefore the optimized coupling is one order of magnitude larger than in the crystal discussed in Sec.~\ref{optomagnonic_crystal_sec}. 
As before we want to gauge this value by comparing it to the analytical estimate given in Eq.~\ref{optomagnonic_coupling_estimate}. 
The optimal coupling in the optimized crystal is $|G_{\text{optimal}}^{\text{F}}| = 2 \pi \times \SI{0.5}{\mega\hertz}$.
Again the magnetic mode volume bounds the coupling due to the smaller size of the magnetic mode compared to the optical mode which also extends to the $\text{Si}_3\text{N}_4$ layers.
This results in a overlap measure (see Eq.~\ref{overlap_measure}) of $\mathcal{O} = 0.04$. 
Therefore the mode overlap is increased by $25\%$ compared to the un-optimized crystal.
Evaluating the directionality measure given in Eq.~\ref{direction_measure} gives $\mathcal{D} = 53 \%$ which is just slightly larger than in the un-optimized case.
Taking both measures into account the analytical coupling estimate shrinks to $|G_{\text{expected}}^{\text{F}}| = \mathcal{O} \cdot \mathcal{D} \cdot |G_{\text{optimal}}^{\text{F}}| \approx 2 \pi \times \SI{10}{\kilo\hertz}$ which lies slightly above the numerically obtained value.
Although the fine structure peaks of the optical spin density and the magnon mode still do not coincide (see Fig.~\ref{figure_11}), the coupling values are improved by ``pulling" the optical and magnetic modes completely into the defect area by the insertion of elliptical holes, creating an overlap area with high density of both modes.

The Cotton-Moutton effect in this structure evaluates to $|G^{\text{C}}_{\text{num}}| = 2 \pi \times \SI{1}{\kilo \hertz}$ and results in a total  coupling of $|G_{\text{num}}| =|G^{\text{F}}_{\text{num}}+G^{\text{C}}_{\text{num}}|= 2 \pi \times \SI{2}{\kilo \hertz}$. We can conclude that in this case both effects interfere constructively for the total coupling.

The cooperativity per photon and magnon in this case is $\mathcal{C}_0^{\text{op}} \sim 2 \cdot 10^{-11}$, which can be enhanced to $\mathcal{C}_{\text{op}} \sim 0.5 \cdot 10^{-6}$ by the number of photons trapped in the cavity.
Thus the cooperativity at maximum photon density is slightly lower as in the crystal presented above, a consequence of the reduced quality factor of the optical mode.

\section{Conclusion} \label{conclusion_sec}

We proposed an optomagnonic crystal consisting of a one-dimensional array with an abrupt defect. We showed that this structure acts as a Bragg mirror both for photon and magnon modes, leading to co-localization of the modes at the defect. By proper design and taking into account the required symmetries of the modes in order to optimize the coupling, we showed that coupling values in the kHz range are possible in these structures. This value is orders of magnitude larger than the experimental state of the art in the field, but still rather small compared to the theoretically predicted optimal value for micron sized structures, which is in the range of $\sim 10^{-1} \SI{}{MHz}$~\cite{optomagnonic_coupling_first}. 

We showed that the strength of the coupling in our proposed structure is still limited largely by the sub-optimal mode overlap, $<5\%$. Further optimization in design is moreover needed in order to boost the cooperativity value, which is limited mainly by the optical losses. The simultaneous optimization is challenging, due to the complexity of the demagnetization fields in patterned geometries. Whereas it is well known that a tapered defect (that is, a smooth defect) can highly increase the optical quality factors, its effect on the magnetic modes is non-trivial and is disadvantageous for localizing the magnon modes of the kind used in this work. Other magnon modes, however, could be explored in this case. More complex geometries, including one-dimensional crystals combining tapering and an abrupt defect, or two-dimensional crystals, are good candidates to be explored in order to improve quality factors and coupling. The first results shown in this work point to the promise of designing the collective excitations in optomagnonic systems via geometry, in order to boost the coupling strength and minimize losses, paving the way for applications in the quantum regime. 

\section{Acknowledgements}
We thank Clinton Potts and Tahereh Parvini for insightful discussions.
J.G. acknowledges financial support from the International Max Planck Research School - Physics of Light (IMPRS-PL). H.H. acknowledges funding from the Deutsche Forschungsgemeinschaft (DFG, German Research Foundation) under Germany's Excellence Strategy EXC-2111-390814868. S.S. and S.V.K. acknowledge funding from the Max Planck Society through an Independent Max Planck Research Group. S.V.K also acknowledges funding from the Deutsche Forschungsgemeinschaft (DFG, German Research Foundation) through Project-ID 429529648 -- TRR 306 QuCoLiMa (``Quantum Cooperativity of Light and Matter").  

\section*{Appendix A: Normalization of magnon modes}
In this section, we discuss the normalization of magnons over a general magnetization texture. The magnetization satisfies the Landau-Lifshitz (LL) equation
\begin{equation}
    \frac{d\boldsymbol{M}}{dt}=-\frac{g\mu_{\text{B}}\mu_{0}}{\hbar}\boldsymbol{M}\times\left(\boldsymbol{H}_{\mathrm{ex}}+\Heff{\boldsymbol{M}}\right),
\end{equation}
where $\boldsymbol{H}_{\mathrm{ex}}$ is an external field and $\boldsymbol{H}_{\mathrm{eff}}$ is a linear functional which can be interpreted as the effective field generated by spin-spin interactions such as exchange, dipolar, etc. Let the static solution, i.e. putting $d\boldsymbol{M}/dt=0$, be $M_{\text{s}}\boldsymbol{m}_{0}(\boldsymbol{r})$ with saturation magnetization $M_{\text{s}}$ and unit vector $\boldsymbol{m}_{0}\cdot\boldsymbol{m}_{0}=1$. This magnetization generates an effective field of the form
\begin{equation}
    \Heff{M_{\text{s}}\boldsymbol{m}_{0}(\boldsymbol{r})}=H_{0}(\boldsymbol{r})\boldsymbol{m}_{0}(\boldsymbol{r})-\boldsymbol{H}_{\mathrm{ex}}(\boldsymbol{r}),
\end{equation}
where the function $H_{0}(\boldsymbol{r})$ depends on the nature of spin-spin interactions. The magnon modes $\delta\boldsymbol{m}_{\gamma}(\boldsymbol{r})e^{-i\omega_{\gamma}t}$ are found by the linearized LL equation
\begin{equation}
    i\omega_{\gamma}\delta\boldsymbol{m}_{\gamma}=\frac{g\mu_{\text{B}}\mu_{0}}{\hbar}\left[\boldsymbol{m}_{0}\times\delta\boldsymbol{h}_{\gamma}+H_{0}\delta\boldsymbol{m}_{\gamma}\times\boldsymbol{m}_{0}\right],\label{eq:LinLL}
\end{equation}
where $\delta\boldsymbol{h}_{\gamma}=\Heff{\delta\boldsymbol{m}_{\gamma}}$.

The LL equation can be derived from the Hamiltonian
\begin{equation}
    \bar{H}=-\mu_{0}\int \text{d} V \left[\frac{\boldsymbol{M}\cdot\Heff{\boldsymbol{M}}}{2}+\boldsymbol{H}_{\mathrm{ex}}\cdot\boldsymbol{M}\right].
\end{equation}
Up to quadratic terms in $\delta\boldsymbol{m}$, we expand the magnetization
\begin{equation}
    \boldsymbol{M}\approx M_{\text{s}}\left(1-\frac{\delta\boldsymbol{m}\cdot\delta\boldsymbol{m}}{2}\right)\boldsymbol{m}_{0}+M_{\text{s}}\delta\boldsymbol{m},\label{eq:Mag}
\end{equation}
and the effective field
\begin{equation}
    \Heff{\boldsymbol{M}}\approx\left(1-\frac{\delta\boldsymbol{m}\cdot\delta\boldsymbol{m}}{2}\right)\left(H_{0}\boldsymbol{m}_{0}-\boldsymbol{H}_{\mathrm{ex}}\right)+\delta\boldsymbol{h},\label{eq:H}
\end{equation}
where
\begin{equation}
    \delta\boldsymbol{A}=\sum_{\gamma}\left[\delta\boldsymbol{A}_{\gamma}\beta_{\gamma}+\delta\boldsymbol{A}_{\gamma}^{*}\beta_{\gamma}^{*}\right],
\end{equation}
with $\boldsymbol{A}$ being $\boldsymbol{m}$ or $\boldsymbol{h}$ and $\beta_{\gamma}$ being magnon amplitudes, i.e. classical counterpart of $\hat{b}_{\gamma}$ defined in Eq. (10) in the main text. The above form ensures $\boldsymbol{M}\cdot\boldsymbol{M}=M_{\text{s}}^{2}$ up to second order in $\delta\boldsymbol{m}$. The Hamiltonian becomes (ignoring a constant term)
\begin{equation}
\begin{split}
    \bar{H}=-\frac{\mu_{0}M_{\text{s}}}{2}\int \text{d} V  \big[&-H_{0}\delta\boldsymbol{m}\cdot\delta\boldsymbol{m} +\delta\boldsymbol{m}\cdot\delta\boldsymbol{h}\\
    &+\delta\boldsymbol{m}\cdot\boldsymbol{H}_{\mathrm{ex}}+\boldsymbol{m}_{0}\cdot\delta\boldsymbol{h} \big].
    \end{split}
\end{equation}
The last two terms are linear in $\delta\boldsymbol{m}_{\gamma}$ and thus should be zero for $\delta\boldsymbol{m}_{\gamma}$ to correspond to a magnon mode. We can simplify the second term by finding the component of $\delta\boldsymbol{h}_{\gamma}$ perpendicular to $\boldsymbol{m}_{0}$ using Eq. (\ref{eq:LinLL}),
\begin{equation}
    \delta\boldsymbol{h}_{\gamma}-\boldsymbol{m}_{0}\left(\boldsymbol{m}_{0}\cdot\delta\boldsymbol{h}_{\gamma}\right)=H_{0}\delta\boldsymbol{m}_{\gamma}-\frac{i\hbar\omega_{\gamma}}{g\mu_{\text{B}}\mu_{0}}\boldsymbol{m}_{0}\times\delta\boldsymbol{m}_{\gamma}.
\end{equation}
Using this and ignoring the linear terms, the Hamiltonian simplifies to
\begin{equation}
\begin{split}
    \bar{H}=\frac{\mu_{0}M_{\text{s}}}{2}\sum_{\mu\gamma}\frac{i\hbar\omega_{\mu}}{g\mu_{\text{B}}\mu_{0}}\int \text{d}V
    \boldsymbol{m}_{0}\cdot\Big[\big(\delta\boldsymbol{m}_{\mu}\beta_{\mu}&-\delta\boldsymbol{m}_{\mu}^{*}\beta_{\mu}^{*}\big) \\
    \times\big(\delta\boldsymbol{m}_{\gamma}\beta_{\gamma}&+\delta\boldsymbol{m}_{\gamma}^{*}\beta_{\gamma}^{*}\big)\Big].
    \end{split}
\end{equation}
As the eigenmodes should diagonalize the Hamiltonian to $\sum\hbar\omega_{\gamma}\left|\beta_{\gamma}\right|^{2}$, we should have
\begin{equation}
    \int dV\boldsymbol{m}_{0}\cdot\left(\delta\boldsymbol{m}_{\gamma}\times\delta\boldsymbol{m}_{\mu}\right)=0,
\end{equation}
and
\begin{equation}
    iM_{\text{s}}\int \text{d}V\boldsymbol{m}_{0}\cdot\left(\delta\boldsymbol{m}_{\mu}\times\delta\boldsymbol{m}_{\gamma}^{*}\right)=g\mu_{\text{B}}\delta_{\mu\gamma}.
\end{equation}
For $\mu=\gamma$, this gives the normalization for magnons. For circularly polarized magnons with $\delta\boldsymbol{m}_{\mu}=\delta m(\boldsymbol{y}+i\boldsymbol{z})/\sqrt{2}$ and $\boldsymbol{m}_{0}=\boldsymbol{x}$, the normalization becomes
\begin{equation}
    \int \text{d}V\delta m^{2}=\frac{g\mu_{\text{B}}}{M_{\text{s}}}.
\end{equation}

\section*{Appendix B: Numerical settings - optical simulations}
In the following we shortly discuss how the optical band structure and the optical modes can be evaluated numerically.
In our work we used two different computational methods: for calculations done in the time domain we use the electromagnetic simulation tool MEEP~\citep{MEEP}, whereas for calculations done in the frequency domain we use the finite element solver Comsol~\cite{comsol}.
We use two simulation tools since with MEEP it is much easier to obtain the band structure of the crystal, and with Comsol the exact mode shape.

\subsubsection{MEEP}
MEEP in general solves for Maxwell's equations in the time domain within some finite composite volume. 
Therefore it essentially performs a kind of numerical experiment~\cite{MEEP}. 
We use MEEP for simulating the band structure of a YIG crystal without defect in order to find its band gap and its corresponding mid-gap frequency. 
Furthermore we use a transmission spectrum simulation to optimize the defect size in order to get the least lossy localized mode. 
For finding the exact frequency of the localized mode we also simulate its spatial shape in the time domain.
For simplicity we simulate the YIG crystal in a 2D model (see~\cite{photonic_crystals_molding_the_flow_of_light, MEEP_tutorial}).
Its material parameters are set by the relative permittivity of $\varepsilon =5$.
In order to account for the leakage of the electromagnetic field the investigated crystal is surronded by a finite size air region large enough so that the leaking electric field decays before it reaches the boundaries, in order to avoid spurious reflection effects. 
This is achieved by choosing the distance between the surface of the crystal and the boundary of the air region as $d_{\text{air}}=3 \cdot \lambda_0$ where $\lambda_0$ is the vacuum wavelength ($d_{\text{air}}=3\SI{3.6}{\micro\metre}$ in our case).
Furthermore, we need boundary conditions along the outside of the air region that are transparent to the leaking field such that the truncated air region represents a reasonable approximation of free space.
Therefore we use a perfectly matched layer at the boundaries of the air region which absorbs all outgoing waves. 
The thickness of this layer should be at least a vacuum wave length ~\cite{comsol_pml_scattering}.
The whole geometry is meshed by one single resolution parameter which discretizes the structure in time and space and gives the number of pixels per distance unit. For all band simulations we used a resolution of $40$ pixels, whereas in case of the transmission spectrum we used a resolution of $20$ pixels and a resolution of $50$ pixels in case of the mode shapes~\cite{MEEP_tutorial}. \\[0.2cm]
$Band~structure~simulations$\\[0.1cm]
For obtaining the band structure we use a YIG crystal without defect ($d=a$) and therefore we can simulate only one single unit cell with a side length of $a$ containing one air hole, and apply an infinite repetition of this cell at each side in $\hat{x}$-direction. 
Since we expect the mid-gap frequency of the crystal with defect to be around $\SI{240}{\tera\hertz}$, we excite the crystal with a gaussian pulse with a center-frequency of $\SI{225}{\tera\hertz}$ and a width of $\SI{450}{\tera\hertz}$ to cover all modes around the band gap. 
We center the pulse peak at an arbitrary postion $(x=0.00123, y=0)$ in order to couple the pulse to an arbitrary mode. 
Since we want to simulate only TE-like modes, in order to save  computational time the pulse only has a $H_z$ component.
For decreasing the computation time even more, we apply an odd mirror symmetry plane for $y=0$. 
The mirror symmetry for $x=0$ is broken by a boundary condition for $0 < k_x < \pi$~\cite{MEEP_tutorial}.
\\[0.2cm]
$Transmission~spectrum~simulations$\\[0.1cm]
For optimizing the defect size we simulate a transmission spectrum for frequencies at the band gap by measuring the flux at the end of the waveguide stemming from a source at the other end.
The measured flux then is normalized to the flux of a waveguide without holes. 
We therefore simulate the transmission spectrum as a function of different defect sizes and use the defect size which gives the highest transmission.
In order to consider only TE-like modes where the electric field lies in plane we need to excite the system with a $J_y$-current source transverse to the propagation direction which is achieved by a gaussian pulse with only a $E_y$-component.  
Its center frequency thereby is $\SI{222}{\tera\hertz}$ (simulated mid-gap frequency) and its width is $\SI{90}{\tera\hertz}$ ($>$ band width). 
Also in this case we apply an odd mirror symmetry for $y=0$ for decreasing the simulation time. 
We note that the mirror symmetry for $x=0$ is broken by the source since it is located at the edge of the waveguide~\cite{MEEP_tutorial}.
\\[0.2cm]
$Mode~shape~simulations$\\[0.1cm]
For evaluating the mode frequency of the localized mode within the band gap we simulate the time evolution of this single mode by exciting it by a gaussian pulse with a center frequency of 
 $\SI{203}{\tera\hertz}$ (frequency of the peak in the transmission spectrum) and a width of $\SI{15}{\tera\hertz}$.
Since in this simulation no symmetry is broken we also apply an odd mirror symmetry for $x=0$ and $y=0$ for obtaining only a TE-like mode~\cite{MEEP_tutorial}.

\subsubsection{Comsol}
We use Comsol to find the spatial mode shape. Therefore we use the ``Electromagnetic waves, Frequency domain" package of COMSOL's ``RF module" which solves for the Helmholtz equation of the form
\begin{equation}
\boldsymbol{\nabla} \times \frac{1}{\mu_{\text{r}}} \left( \boldsymbol{\nabla} \times \boldsymbol{E} \right) - k_0^2 \left( \varepsilon_{\text{r}} - \frac{i \sigma}{\omega \varepsilon} \right) \boldsymbol{E} = 0,
\end{equation}
where $k_0$ indicates the vacuum wave number, $\omega$ the angular frequency, $\mu_r$ the relative permeability and $\varepsilon_0$ the vacuum permittivity. 
Contrary to the MEEP simulations above, we simulate the full geometry composite of a YIG layer sandwiched by two $\text{Si}_3\text{N}_4$ layers. 
The used material parameters thereby are $\varepsilon_{\text{YIG}} = 5$, $\varepsilon_{\text{Si}} = 4$, $\mu_{\text{YIG}} = \mu_{\text{Si}}= 0$, and $\sigma_{\text{YIG}} = \sigma_{\text{Si}} = 0$ with $\mu$ the relative permeabilty and $\sigma$ the conductivity.
Again we also need to simulate an truncated air region around the crystal which is able to absorb the outgoing radiation.  
The corresponding material parameters are $\varepsilon_{\text{air}} = \mu_{\text{air}} = 1$ and $\sigma_{\text{air}} = 0$.
Besides perfectly matched layers we also can use second order scattering boundary conditions at the air surfaces given by the expression \cite{comsol_pml_scattering}
\begin{equation}
\boldsymbol{n} \cdot \nabla E_z + i k_0 E_z - \frac{i}{2 k_0} \nabla_t^2 E_z = 0
\end{equation}
with $\boldsymbol{n}$ the normal vector to the considered plane. 
For large enough air regions both approaches are almost equivalent as long as the leaking field is propagating normal to the air surfaces.
In order to account for a large enough air region we choose the distance between the surfaces of the crystal and the air boundaries as $\SI{4.5}{\micro\metre}$.
For reducing the simulation time we use the symmetry requirements of a TE-like mode. 
Therefore, we cut the geometry into an eighth of the whole structure and apply perfect electric conductor boundary conditions ($\boldsymbol{n} \times \boldsymbol{E} = 0$) at the cut surfaces along $x = 0$ and $y = 0$ and a perfect magnetic conductor boundary condition ($\boldsymbol{n} \times \boldsymbol{H} = 0$) at the cut surface along $z=0$. 
The full solution is then obtained by using the symmetry requirements of a TE-like mode.
The whole geometry is meshed by a physics-controlled tetrahedral mesh with a maximum element size of $\lambda_0/5 \approx \SI{0.3}{\micro\metre}$~\cite{comsol_mesh}.
We note that in case of a physics-controlled mesh Comsol automatically meshes the material areas of interest with a finer mesh and uses a coarser mesh e.g. for the air regions.

\section*{Appendix C: Numerical settings - magnetic simulations}
In this section we briefly discuss how the magnetic band structure and magnetic mode shape is obtained numerically.
For evaluating the magnetization dynamics we use the finite difference tool MuMax3~\cite{mumax} which solves for the Landau-Lifshitz-Gilbert equation of the form
\begin{equation}
\frac{\partial \boldsymbol{m}}{\partial t} = g \frac{1}{1+\alpha^2} \left[\boldsymbol{m} \times \boldsymbol{B}_{\text{eff}} + \alpha (\boldsymbol{m} \times ( \boldsymbol{m} \times \boldsymbol{B}_{\text{eff}}))\right]
\end{equation}
with $\boldsymbol{m} = \boldsymbol{M}/M_{\text{s}}$ the local reduced magnetization of one simulation cell, $g$ the gyromagnetic ratio, $\alpha$ the damping parameter, and $\boldsymbol{B}_{\text{eff}}$ an effective field which contributions can be found in~\cite{mumax}.
As material parameters we used the parameters for YIG, $M_{\text{s}} = \SI{140}{\kilo\ampere/\metre}$ (saturation magnetization), $A_{\text{ex}} = \SI{2}{\pico\joule/\metre}$ (exchange constant), $K_{\text{c}1} = \SI{-610}{\joule/\metre^3}$ (anisotropy constant) with the anisotropy axis along $\hat{z}$.
In order to accelerate the simulations, we used an increased Gilbert Damping parameter  $ \alpha = 0.008$ (compare to $\alpha\approx 10^-5$ for YIG)~\cite{yig_parameters}.
The used meshgrid had $(1024,~50,~5)$ cells in the $(\hat{x},~\hat{y}, \hat{z})$-direction what guarantees to take the exchange interaction into account ($l_{\text{\text{ex}}} \approx \SI{13}{\nano\metre}$ for YIG).

In general, in all our simulations the spin wave dynamics is excited via an external pulse field and the time evolution  is recorded for all three magnetization components.
For post processing the output of the form $m_i(x,y,z)$ with $i = (x,y,z)$ saved for all simulated time steps separately we create for each magnetization component $i = (x,y,z)$ a 4D-array of the form $m_i (t, x, y, z)$.
\\[0.2cm]
$Band~structure~simulations$\\[0.1cm]
In order to obtain the band structure along a specific direction $j$ with $j = (x,y,z)$ e.g. chosen to be the $\hat{x}$-direction, we reduce the four dimensional array to a two dimensional array of the form $m_ i (t,x) = \sum_m^{n_y} \sum_n^{n_z}\delta m_i(t, x, y_m, z_n)$ and perform a 2D Fourier transform on this array $\delta m_i (f, k_x) = \mathcal{FT}_{2\text{D}} [m_ i (t,x)]$ resulting in the band diagram along the chosen direction.
For increasing the resolution in the band diagram we plot the quantity $\sqrt{|\delta m_ i (t,x)|}/max(|\delta m_ i (t,x)|)$.
\\[0.2cm]
$Mode~shape~simulations$\\[0.1cm]
In order to obtain the mode shape we perform a space-dependent Fourier transform in time on each array entry separately  $\delta m_i (f, x, y, z) = \mathcal{FT}_{1\text{D}} [m_ i (t,x,y,z)]$.

\section*{Appendix D: Evaluation of the mode volumes}
For evaluating the mode volume numerically we first, due to numerical errors, need to identify all cells of the simulated array (either containing $\delta \boldsymbol{m} (\boldsymbol{r})$ in case of the magnetic mode volume or $\boldsymbol{E}(\boldsymbol{r})$ in case of the optical mode volume) which contribute to the volume by a high enough mode density. This means we need to define a threshold which determines if a cell should contribute to the mode volume or not.
We define this threshold by
\begin{equation}
    \mathcal{T} = \frac{\text{max}(|\boldsymbol{x}|) - \text{min}(|\boldsymbol{x}|)}{5},
    \label{vol_count}
\end{equation}
where $\boldsymbol{x}$ is the content of the cell (either $\boldsymbol{x} = \delta \boldsymbol{m}$ or $\boldsymbol{x} = \boldsymbol{E}$).
For being able to count the cells which contribute to the volume we create an additional array matching the simulated arrays in size.
This array then contains ones if the absolute value of the cell ($|\boldsymbol{x}|$) in the original array is larger than the threshold given in Eq.~\eqref{vol_count} and zeros if $|\boldsymbol{x}|$ is smaller.
The mode volume is then obtained by summing over this array (giving the number of cells contributing to the volume) and by multiplying this such calculated number by the volume of one cell $s_x \cdot s_y \cdot s_z$.

For evaluating the overlap volume between the magnetic mode ($x = \delta \boldsymbol{m}$) and in this case the optical spin density ($\boldsymbol{E}^* \times \boldsymbol{E}$) we create the same additional arrays as above identifying the cells which contribute to their corresponding mode volume.
For identifying the cells where the magnon mode and the optical spin density overlap we create a third array again matching the size of the original array. 
But this array now contains ones if the corresponding cells of the ``threshold arrays" both in the magnetic and optical case contain a one, otherwise we set the cell value to zero. 
The overlap volume is then obtained by summing over this third array (giving the number of cells contributing to the overlap) and by multiplying this such calculated number by the volume of one cell $s_x \cdot s_y \cdot s_z$.

\bibliography{references}
\end{document}